\documentclass[11pt,a4paper]{article}
\usepackage{jcappub}
\usepackage{epsf,epstopdf}

\title{Fermi-LAT constraints on dark matter annihilation cross section
from observations of the Fornax cluster}
\author[a,b]{Shin'ichiro Ando}
\author[c,d,e]{and Daisuke Nagai}
\affiliation[a]{Institute for Theoretical Physics, University of
Amsterdam, 1090 GL Amsterdam, The Netherlands}
\affiliation[b]{Gravitation and Astroparticle Physics, University of
Amsterdam, 1090 GL Amsterdam, The Netherlands}
\affiliation[c]{Department of Physics, Yale University, New Haven, CT 06520, USA}
\affiliation[d]{Department of Astronomy, Yale University, New Haven, CT 06520, USA}
\affiliation[e]{Yale Center for Astronomy and Astrophysics, Yale University, New Haven, CT 06520, USA}
\emailAdd{s.ando@uva.nl}
\emailAdd{daisuke.nagai@yale.edu}
\abstract{
We analyze 2.8-yr data of 1--100 GeV photons for clusters of galaxies,
collected with the Large Area Telescope 
onboard the Fermi satellite. By analyzing 49 nearby massive clusters located at 
high Galactic latitudes, we find no excess gamma-ray emission towards 
directions of the galaxy clusters.  Using flux upper limits, we show that the Fornax 
cluster provides the most stringent constraints on the dark matter
annihilation cross section. 
Stacking a large sample of nearby clusters does not help improve the limit for most dark matter models.
This suggests that a detailed modeling of the Fornax cluster is important for setting
robust limits on the dark matter annihilation cross section based on clusters. We 
therefore perform the detailed mass modeling and predict the expected dark matter annihilation 
signals from the Fornax cluster, by taking into account effects of dark matter contraction and 
substructures.
By modeling the mass distribution of baryons (stars and gas) around a central bright 
elliptical galaxy, NGC~1399, and using a modified contraction model
motivated by numerical simulations,
we show that the
dark matter contraction boosts the annihilation signatures by a factor of 4.  For dark 
matter masses around 10\,GeV, the upper limit obtained on the annihilation cross section times 
relative velocity is
$\langle \sigma v \rangle \lesssim (2$--$3)\times 10^{-25} \, \mathrm{cm^3 \, s^{-1}}$, 
which is within a factor of 10 from the value required to explain the 
dark matter relic density. This effect is more robust than the annihilation
boost due to substructure, and it is more important unless the mass of the
smallest subhalos is much smaller than that of the Sun.
}
\arxivnumber{1201.0753}
\begin{document}
\maketitle
\flushbottom

\section{Introduction}
\label{sec:intro}

Revealing the nature of dark matter is one of the important and
challenging goals of modern physics, astrophysics, and cosmology.
One of the avenues involves searching for annihilation signatures from
dark matter halos in a form of gamma rays, neutrinos, and charged particles.
Such a possibility is theoretically well motivated especially if dark
matter is made of weakly interacting massive particles (WIMPs) such as
the supersymmetric neutralino~\cite{Jungman1996}.
In addition, in order to thermally produce dark matter at the right
abundance in the present Universe, WIMPs have to have some certain
annihilation cross section (times relative velocity), most preferably,
$\langle \sigma v \rangle = 3 \times 10^{-26}\, {\rm cm}^3\, {\rm
s}^{-1}$~\cite{Jungman1996, Bergstrom2000, Bertone2005}.
The recently launched Fermi Gamma-Ray Space Telescope and ground-based
atmospheric \v Cerenkov telescopes such as HESS, MAGIC, and VERITAS have
started to provide interesting constraints on the annihilation cross section of
dark matter particles from various astrophysical objects, ranging from
dwarf galaxies to the isotropic diffuse gamma-ray background, i.e., the
emission coming from the entire Universe.

Clusters of galaxies are the largest concentrations of dark matter pulled together
by gravity, and therefore, an attractive astrophysical object to look for the 
annihilation signals.  The Fermi Large Area Telescope (LAT) 
collaboration analyzed their 11-month data to place constraints on the 
dark matter annihilation cross section~\cite{FermiClusterDM} and
decay~\cite{Dugger2010} using observations of several nearby galaxy
clusters.
The obtained upper limits have already started to exclude some of the 
parameter space of WIMPs, and these cluster-based constraints are quite 
comparable to those obtained from dwarf 
galaxies~\cite{FermiDwarfDM, FermiDwarfDM2, FermiDwarfDM3} 
and the diffuse gamma-ray background~\cite{FermiDiffuseDM}.

In the previous analyses as well as theoretical studies on dark matter
annihilation in galaxy clusters, it is commonly assumed that the dark 
matter density profile is smooth and characterized by the Navarro-Frenk-White 
(NFW) form~\cite{NFW}. However, it has been pointed out that the effects 
of subhalos might significantly boost the dark matter annihilation signal 
%by a factor of $\sim$1000 
for massive clusters~\cite{SanchezConde2011, Pinzke2011, Gao2011}, if the
smallest subhalo masses are of around Earth size as implied by arguments
of the kinetic decoupling of WIMP particles in the early
Universe~\cite{Green2004, Green2005, Diemand2005, Diemand2006,
Profumo2006}.
In the case of the smooth NFW-like halos, the current upper limits on the 
cross section is about $\sim$30 times larger than the canonical value ($\langle \sigma v
\rangle = 3 \times 10^{-26}\, {\rm cm}^3\, {\rm s}^{-1}$) for the
low-mass dark matter particles around $\sim$10\,GeV.
In the case of Earth-mass substructures, on the other hand, the
canonical cross section has already been excluded for the low-mass
WIMPs~\cite{FermiClusterDM, Huang2011}, although there is a large
uncertainty on the subhalo models, especially since it involves 
significant extrapolation of the subhalo mass functions down to 
small scales. 

There is another effect that {\it must be taken into account but has
received little attention} to date---the effect of dark matter contraction 
due to baryonic infall.
Unlike dark matter, baryons can lose energy via radiative cooling
and they fall towards the center of the halo, and as the result, modify 
gravitational potential in the central region.
Dark matter is then dragged towards the center because of the deepened
potential, modifying the density profile of dark matter \cite{Blumenthal1986, 
Ryden1987, Gnedin2004,Gnedin2011}. Its implications for dark 
matter annihilation have been discussed in the context of galaxy-size
halos~\cite{Prada2004, Gustafsson2006}, but not in the context of cluster-size
halos. Galaxy clusters contain a large amount of baryonic gas and
stars, and in fact, the stars dominate the gravitational potential at
central regions.  Although there is no firm observational evidence for 
or against the halo contraction in massive clusters at the moment, the 
effect of the dark matter contraction is likely important for cluster-size halos.

In this paper, we analyze Fermi-LAT 2.8-yr data in order to constrain the dark 
matter annihilation signatures from galaxy clusters. As we shall show in this work,
among several clusters analyzed thus far, the most stringent constraint on 
the annihilation cross section is obtained with Fornax, a nearby, well 
measured galaxy cluster located at 20\,Mpc away from Earth. 
We therefore focus on interpreting observations of the Fornax cluster, 
paying particular attention to modeling of dark matter contraction due
to baryon dissipation as well as substructure boost, and address the
relative importance of these two effects.  
Apart from its proximity and its large mass ($\sim$10$^{14}\, h^{-1}\,M_\odot$), 
there are several reasons why the Fornax cluster is ideal for detailed modeling. 
First, the Fornax cluster does not host any bright active galactic
nuclei, unlike the Virgo and Perseus clusters. Secondly, it has regular
thermal gas profiles and 
a spherical central massive elliptical galaxy, NGC~1399, and this feature makes 
the calculation of contraction based on the assumption of spherical symmetry 
quite reasonable. 
In order to infer the distribution of baryons, we use high-resolution observations 
of surface brightness profiles with Hubble Space Telescope (HST) for stars 
and with ROSAT X-ray Telescope for thermal gas.  Using the well-measured 
stellar and gas profiles, we then compute the dark matter density
profile taking into account the dark matter contraction. 

%To make this paper self-contained, we also analyze the Fermi-LAT 2.8-yr
%data sample and obtain the upper limits on annihilation cross section
%for various models of dark matter contraction.
%As a result, we find that this effect will strengthen the cross-section
%upper bound by a factor of 4 with uncertainty range of 2--6, which
%corresponds to parameter ranges compatible with the results of numerical
%simulations~\cite{Gnedin2011}.
%We then compare this with results of dark matter substructure, and
%conclude that the former is much more robust than the latter.
%In addition, we also show that unless the minimum subhalo mass is much
%smaller than the solar mass, the effects of baryons cannot be ignored.

This paper is organized as follows. In Sec.~\ref{sec:dark matter profiles}, we 
present the formulation to compute intensities of gamma rays from dark matter 
annihilation, and models of dark matter density profiles, halo contraction 
due to baryon dissipation, and substructures from numerical simulations.
In Sec.~\ref{sec:sample}, we discuss the results of analysis for a large
sample of galaxy clusters, including the upper limits on individual
clusters and stacking analysis, highlighting that a detailed
mass modeling of the Fornax cluster is indeed more important than the 
statistical constraint. Procedures for the data analysis of the Fermi-LAT data 
are summarized in Appendix~\ref{sec:analysis}. 
In Sec.~\ref{sec:Fornax}, we show results for the Fornax cluster, in which 
we present a detailed mass modeling of dark matter, stellar, and gaseous 
components in Fornax and discuss the relative importance of halo contraction
and subhalo boost effects in modeling the annihilation signal from 
clusters.
We finally conclude the present paper in Sec.~\ref{sec:conclusions}.
Throughout the paper, we assume the flat cold dark matter cosmology with
cosmological constant ($\Lambda$CDM), and adopt $\Omega_m = 0.277$,
$\Omega_\Lambda = 0.723$, $H_0 = 100\, h \, \mathrm{km \, s^{-1}
\, Mpc^{-1}}$, and $h = 0.7$.

\section{Gamma-ray intensity and dark matter density profiles}
\label{sec:dark matter profiles}

In order to compute the gamma-ray intensity from dark matter
annihilation, one needs to specify the dark matter annihilation 
mechanisms and the dark matter distribution in clusters. 
Here, after briefly introducing formalism to compute the intensity in
Sec.~\ref{sub:formulation}, we discuss the density profiles obtained 
with numerical N-body simulations in the literature, where gravitational 
structure formation and evolution of only dark matter particles are followed.
We introduce the smoothly distributed large-scale halo component
(Sec.~\ref{sub:DM}), effects of halo contraction due to baryon dissipation
(Sec.~\ref{sub:contra}), and small-scale substructure (Sec.~\ref{sub:subhalo}).
%We present results for the Fornax cluster in Sec.~\ref{sec:Fornax} and
%the large statistical sample in Sec.~\ref{sec:sample}.

\subsection{Gamma-ray intensity from dark matter annihilation}
\label{sub:formulation}

The gamma-ray intensity, here defined as the number of photons received per
unit area, time, solid angle, and energy, i.e., $I_\gamma \equiv
dN_\gamma / dAdtd\Omega dE$, from annihilation of dark matter particles
$\chi$, in a halo is given by
\begin{equation}
 I_\gamma(E,\theta) =
  \frac{\langle \sigma v \rangle}{8\pi (1+z)^2 m_\chi^2}
  \left.\frac{dN_{\gamma,{\rm ann}}}{dE^\prime}
  \right|_{E^\prime = (1+z) E}
  \int ds\ \rho_\chi^2 [r(s,\theta)],
  \label{eq:intensity}
\end{equation}
where $\theta$ is the angle from the halo center, $z$ is the redshift of 
the halo, $m_\chi$ is dark matter mass, and $dN_{\gamma, {\rm ann}}/ dE$
is the gamma-ray spectrum per annihilation.
The intensity is proportional to the line-of-sight ($s$) integral of dark
matter density squared, $\rho_\chi^2$.
%Thus, it is essential to have precise knowledge of the density profile
%of dark matter, which is discussed below in detail.
Here, the density is assumed to be spherically symmetric, and the radius
$r$ is related to $s$ and $\theta$ through $r = \sqrt{s^2 + d_A^2
\theta^2}$, where $d_A$ is the angular diameter distance to the halo.

\begin{figure}
\begin{center}
\includegraphics[width=7.0cm]{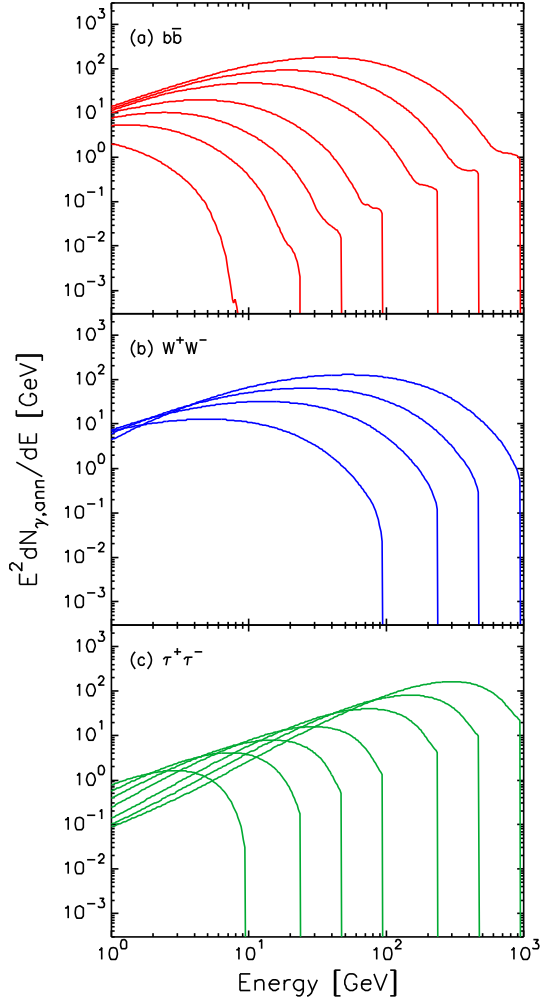}
\caption{Gamma-ray spectrum per dark matter annihilation, $E^2
 dN_{\gamma, {\rm ann}} / dE$ for (a) $b\bar b$, (b) $W^+W^-$, (c)
 $\tau^+\tau^-$ annihilation channels. Assumed dark matter masses are
 $m_\chi = 10$, 25, 50, 100, 250, 500, and 1000 GeV for the panels (a)
 and (c), and the same unless $m_\chi < m_W = 80.4$ GeV for the panel
 (b).}
\label{fig:DM_spectrum}
\end{center}
\end{figure}

The gamma-ray spectrum depends on the final state of annihilation.
In this study, we adopt three different annihilation channels, $\chi
\chi \to b\bar b$, $W^+ W^-$, and $\tau^+ \tau^-$.
Figure~\ref{fig:DM_spectrum} shows $E^2 dN_{\gamma,{\rm ann}} / dE$
for various dark matter masses between 10~GeV and 1~TeV.
The annihilation spectra for neutralino WIMPs are typically given by
some combination of these three spectra.
Note that there is no annihilation happening if the final state is $W^+
W^-$ and dark matter mass is smaller than $W^\pm$ mass (80.4~GeV).

\subsection{Dark matter density profile: smooth component}
\label{sub:DM}

For the smooth component of the dark matter density profile, $\rho(r)$,
we consider two models: one is the NFW profile~\cite{NFW}, and the other
is the Einasto profile with a central core~\cite{Merritt2006,
Navarro2008}.
The NFW profile is given by
\begin{equation}
 \rho(r) = \frac{\rho_s}{(r/r_s)(r/r_s + 1)^2},
\end{equation}
where $\rho_s$ and $r_s$ are the scale density and scale radius,
respectively.
The virial mass $M_{\rm vir}$ and redshift $z$ are given inputs, and
the virial radius $r_{\rm vir}$ is obtained by solving $M_{\rm vir} =
4\pi r_{\rm vir}^3 \Delta_{\rm vir}(z) \rho_c(z) / 3$, where $\rho_c(z)$
is the critical density, $\Delta_{\rm vir} (z) = 18\pi^2
+ 82 d - 39 d^2$, and $d = \Omega_m (1+z)^3 / [\Omega_m (1+z)^3 +
\Omega_\Lambda] - 1$~\cite{Bryan1998}.
The scale radius $r_s$ is then defined as $r_s = r_{\rm vir} / c_{\rm
vir}$, where $c_{\rm vir}$ is the concentration parameter.
We adopt the result of the mass-concentration relations from
Ref.~\cite{Duffy2008} (see also Refs.~\cite{Neto2007, Gao2008, Zhao2009,
Prada2011}):
\begin{equation}
 c_{\rm vir}(M_{\rm vir},z) = \frac{7.85}{(1+z)^{0.71}}
  \left(\frac{M_{\rm vir}}{2\times 10^{12}\,h^{-1}
   M_\odot}\right)^{-0.081}.
  \label{eq:concentration}
\end{equation}
By taking volume integral of $\rho(r)$ up to the virial radius
$r_{\rm vir}$ and then equating it to $M_{\rm vir}$, we obtain
\begin{equation}
 \rho_s = \frac{M_{\rm vir}}{4\pi r_s^3}
  \left[\ln (1+c_{\rm vir}) - \frac{c_{\rm vir}}{1+c_{\rm
   vir}}\right]^{-1}.
\end{equation}

The Einasto profile, on the other hand, has the following functional
form:
\begin{equation}
 \rho(r) = \rho_e \exp
  \left\{-d_n \left[\left(\frac{r}{r_e}\right)^{1/n}-1\right]\right\},
  \label{eq:Einasto}
\end{equation}
where $n$ and $r_e$ are the free parameters, and $d_n=3n - 1/3 +
0.0079/n$.
The normalization $\rho_e$ is obtained such that the total mass
within $r_{\rm vir}$ agrees with the measured $M_{\rm vir}$:
\begin{equation}
 \rho_e = \frac{M_{\rm vir}d_n^{3n}}{4\pi r_e^3n e^{d_n}}
  \Gamma\left(3n,d_n
	 \left[\frac{r_{\rm vir}}{r_e}\right]^{1/n}\right)^{-1},
\end{equation}
where $\Gamma(a,x)$ is the lower incomplete gamma function.

Note that we also use other definitions of mass, including $M_{\Delta}$
with $\Delta = 200$ or 500, defined as the enclosed mass within a sphere 
with radius $r_{\Delta}$, within which the average density is $\Delta$ times 
the critical density of the universe.
Conversion between different mass definitions is straightforward if one
assumes density profile such as NFW as well as the concentration-mass
relation (e.g.,~\cite{Hu2003}).

\subsection{Dark matter contraction due to baryon dissipation}
\label{sub:contra}

The dark matter density profile is modified as a result of baryon dissipation.  
A so-called dark matter contraction model is based on the conservation of 
the angular momentum:
\begin{equation}
 M_i(\bar r_i) r_i = [M_{{\rm dm}, i}(\bar r_i) + M_{b,f}(\bar r_f)]
  r_f,
  \label{eq:contra}
\end{equation}
where $M(r)$ is the mass enclosed within the radius $r$, subscripts dm
and $b$ are for dark matter and baryons, respectively, and subscripts
$i$ and $f$ represent quantities before and after the contraction,
respectively.
The masses inferred from, e.g., gravitational lensing or X-ray observations 
are the {\it total} mass, and therefore one must assume underlying distributions 
of both baryons and dark matter in order to infer $M_{{\rm dm}, i}(r)$.
Here, when we discuss the {\it initial} density profile (before
contraction), we simply assume that dark matter and baryons trace each
other exactly; namely, $\rho_{{\rm dm}, i}(r) = (1-f_b) \rho_i(r)$ 
and $M_{{\rm dm}, i}(r) = (1-f_b) M_i(r)$, where $f_b$ is the baryon fraction at
virial radius.
For the {\it final} profiles of baryons, on the other hand, we use the observed distribution
of stars and gas  (as discussed in the later section), giving the 
baryon mass distribution to be $M_{b,f}(r) \equiv M_\star (r) + M_{\rm gas}(r)$. 
Finally, $\bar r$ is the orbit-averaged radius for a particle found at
$r$ in a simulation~\cite{Gnedin2004}.
Equation~(\ref{eq:contra}) is solved for $r_f$, a contracted radius, as
a function of the initial radius $r_i$.
The dark matter mass profile after the contraction is then obtained by
$M_{{\rm dm}, f} (r_f) = M_{{\rm dm},i}(r_i(r_f))$.

If particle orbits can be approximated to be circular and if the
contraction happens adiabatically, then we simply have $\bar r = r$.
This is a so-called standard adiabatic contraction model originally
proposed in Ref.~\cite{Blumenthal1986}.
Reference~\cite{Gnedin2004}, based on hydrodynamical 
simulations, presented the modified formula (\ref{eq:contra}) and
pointed out that corrections to the adiabatic contraction model can be
accommodated by modifying a relation between $\bar r$ and $r$ as $\bar
r/r_{\rm vir} = 0.85 (r/r_{\rm vir})^{0.8}$.
This was further investigated by various groups both
theoretically~\cite{Sellwood2005, Choi2006, Colin2006, Gustafsson2006,
Johansson2009, Abadi2010, Duffy2010, Tissera2010} and
observationally~\cite{Biviano2006, Shulz2010, Auger2010, Sonnenfeld2011}.
Recently, Ref.~\cite{Gnedin2011} systematically studied this effect, and
found that the relation
\begin{equation}
\frac{\bar r}{r_0} =
 A_0 \left(\frac{r}{r_0}\right)^w ,
 \label{eq:rbar}
\end{equation}
well represents the simulation results, where $r_0 = 0.03 r_{\rm vir}$,
and $w$ ranges from 0.5 to 1.3, if $A_0$ is fixed to 1.6.
Here we adopt $w = 0.8$ as a canonical value for this parameter, but
also consider the cases of $w = 0.6$ and 1.
In addition, we adopt $A_0 = 1$ and $w = 1$, the standard adiabatic
contraction, as an extreme case scenario.
Equation~(\ref{eq:contra}) has been calibrated only down to $r \sim
10^{-3} r_{\rm vir}$ because of the resolution limit~\cite{Gnedin2011},
but we shall extrapolate this relation further to very small radii.
Although we assume angular momentum conservation, there is no
guarantee that Eq.~(\ref{eq:rbar}) remains valid below the resolution 
limit of the simulations. 
We therefore adopt yet another relation as a conservative scenario
of the contraction: $\bar r = \max [ A_0 r_0 (r/r_0)^w, 10^{-2} r_{\rm
vir}]$, which does not allow $\bar r$ to reach below $10^{-2} r_{\rm vir}$, 
the current resolution limit of simulations. 
Note that in this model the magnitude of contraction is smaller for 
the larger $\bar r$. We discuss the application of this model to the 
Fornax cluster in Sec.~\ref{sec:Fornax}.

\subsection{Effects of substructure}
\label{sub:subhalo}

If galaxy clusters consist of numerous subhalos as implied by numerical
simulations, then the gamma-ray radiation may be dominated by the contribution 
from subhalos, even though they make up a small fraction of the total cluster mass. 
%In this case, the density profile $\rho(r)$ would be much more
%complicated than the smooth NFW or Einasto profile at small scales.
Here, we consider the results of recent numerical simulations of
cluster-size halos that contain subhalos of $M_{\rm sub} > 5\times 10^7
M_\odot$, and its extrapolation down to Earth mass, 
$M_{\rm min} = 10^{-6} M_\odot$, or even smaller mass 
$10^{-12} M_\odot$~\cite{Gao2011}.
(This is because the minimum mass of subhalos for neutralino dark matter
models ranges widely, $M_{\rm min} = 10^{-12}$--$10^{-4}
M_\odot$~\cite{Profumo2006}.)
According to this model, the surface brightness profile contributed by 
subhalos are given by
\begin{eqnarray}
 I_{\gamma,\rm sub}(E,\theta) &=&
  \frac{b(M_{200}) F_{\gamma, {\rm host}}(E)}{\pi\ln (17)}
  \frac{1}{\theta^2 + (\theta_{200}/4)^2},
  \\
 b(M_{200}) &=& 1.6\times 10^{-3}
  \left(\frac{M_{200}}{M_\odot}\right)^{0.39},
\end{eqnarray}
for $\theta \le \theta_{200}$ and for $M_{\rm min} = 10^{-6} M_\odot$.
Here $\theta_{200} \equiv r_{200} / d_A$, the flux $F_{\gamma}$ is
obtained by integrating $I_{\gamma}$ [Eq.~(\ref{eq:intensity})] over the
solid angle, $b = F_{\gamma, {\rm sub}} / F_{\gamma, {\rm host}}$ is the
boost factor, and the subscripts ``sub'' and ``host'' indicate subhalos and 
the host halo, respectively. Note that the boost factor decreases significantly 
for the larger values of minimum subhalo mass $M_{\rm min}$, and it 
scales roughly as $b \propto M_{\rm min}^{-0.2}$~\cite{Gao2011}.

\section{Constraints from a large statistical sample of clusters}
\label{sec:sample}

We discuss constraints on annihilation cross sections from 49 nearby
rich galaxy clusters.
After computing the intensity $I_\gamma (E, \theta)$ with
Eq.~(\ref{eq:intensity}) for these clusters, we analyze the Fermi-LAT 
data for 2.8 years to place constraints on $\langle \sigma v \rangle$.
%Since the angular size of the scale radius, 
%$r_s / d_A$, where most of the annihilation photons are generated in the case 
%of the NFW or Einasto profiles, is generally larger than the angular resolution 
%for the high-energy photons (e.g., at 10\,GeV), it is important to take the
%source extension into account in the analysis.
Unlike the previous study where clusters were assumed to be point-like
sources~\cite{FermiClusterDM}, here we take the extension of the cluster
emission fully into account. Analysis details can be found in the 
Appendix~\ref{sec:analysis}. In this section, we consider only the NFW 
smooth halo model (Sec.~\ref{sub:DM}) for a density profile, and other 
models are discussed in the following section.
%Procedures for data analyses of Fermi-LAT data for 2.8 years and how one
%constrains annihilation cross section are summarized in
%Appendix~\ref{sec:analysis}.

\subsection{Upper limits on individual clusters}
%\label{appendix2}

\begin{table}
\begin{center}
\scriptsize
\caption{Parameters of galaxy clusters analyzed for dark matter
 annihilation. Upper sample is from Chandra Cluster Cosmology
 Project~\cite{Vikhlinin2009a} and lower one is for the X-ray-selected
 bright clusters~\cite{Reiprich2002}. A2199 and A3571 are found in the
 both samples.}
\label{table:clusters}
\begin{tabular}{lccccccc}
\hline \hline 
 & & & & $M_{\rm vir}$ & $r_{\rm vir}$ & $F_{\rm lim}(>0.1\,{\rm GeV})$ & \\ 
Cluster & $l$ ($^\circ$) & $b$ ($^\circ$) & $z$ & ($10^{14} h^{-1} M_{\odot}$) & ($h^{-1}$\,Mpc) & ($10^{-9}$\,cm$^{-2}$\,s$^{-1}$) & $F_{\rm DM}/F_{\rm lim}$ \\ 
\hline 
     A2142 & $ 44.23$ & $ 48.69$ & $0.0904$ & $16.86$ & $2.33$ & $1.93$ & $ 0.15$ \\ 
     A3266 & $272.09$ & $-40.17$ & $0.0602$ & $12.57$ & $2.15$ & $2.23$ & $ 0.23$ \\ 
     A2029 & $  6.50$ & $ 50.55$ & $0.0779$ & $12.05$ & $2.10$ & $4.39$ & $ 0.07$ \\ 
      A401 & $164.18$ & $-38.87$ & $0.0743$ & $12.04$ & $2.10$ & $1.42$ & $ 0.22$ \\ 
      A754 & $239.25$ & $ 24.76$ & $0.0542$ & $11.81$ & $2.12$ & $4.43$ & $ 0.13$ \\ 
      A478 & $182.41$ & $-28.30$ & $0.0881$ & $11.35$ & $2.05$ & $2.06$ & $ 0.10$ \\ 
     A2256 & $111.10$ & $ 31.74$ & $0.0581$ & $10.90$ & $2.06$ & $0.63$ & $ 0.77$ \\ 
     A3667 & $340.85$ & $-33.40$ & $0.0557$ & $10.20$ & $2.01$ & $4.56$ & $ 0.11$ \\ 
      A399 & $164.37$ & $-39.47$ & $0.0713$ & $ 8.53$ & $1.88$ & $3.83$ & $ 0.07$ \\ 
       A85 & $115.05$ & $-72.06$ & $0.0557$ & $ 8.24$ & $1.87$ & $2.53$ & $ 0.16$ \\ 
     A3571 & $316.32$ & $ 28.55$ & $0.0386$ & $ 8.13$ & $1.88$ & $1.19$ & $ 0.72$ \\ 
  ZwCl1215 & $282.50$ & $ 65.19$ & $0.0767$ & $ 7.92$ & $1.83$ & $1.14$ & $ 0.18$ \\ 
     A1795 & $ 33.79$ & $ 77.16$ & $0.0622$ & $ 7.51$ & $1.81$ & $1.01$ & $ 0.30$ \\ 
     A2065 & $ 42.88$ & $ 56.56$ & $0.0723$ & $ 6.83$ & $1.74$ & $1.91$ & $ 0.11$ \\ 
     A3558 & $311.98$ & $ 30.74$ & $0.0469$ & $ 6.54$ & $1.74$ & $0.94$ & $ 0.51$ \\ 
      A119 & $125.70$ & $-64.10$ & $0.0445$ & $ 6.15$ & $1.71$ & $2.57$ & $ 0.20$ \\ 
     A1644 & $304.91$ & $ 45.50$ & $0.0475$ & $ 5.74$ & $1.67$ & $0.90$ & $ 0.46$ \\ 
     A3158 & $265.07$ & $-48.97$ & $0.0583$ & $ 5.63$ & $1.65$ & $1.97$ & $ 0.14$ \\ 
     A3391 & $262.36$ & $-25.16$ & $0.0551$ & $ 5.53$ & $1.64$ & $3.55$ & $ 0.08$ \\ 
     A3562 & $313.31$ & $ 30.35$ & $0.0489$ & $ 4.44$ & $1.53$ & $0.76$ & $ 0.41$ \\ 
     A2147 & $ 28.81$ & $ 44.49$ & $0.0355$ & $ 4.19$ & $1.51$ & $3.25$ & $ 0.18$ \\ 
     A3376 & $246.53$ & $-26.29$ & $0.0455$ & $ 4.06$ & $1.49$ & $1.61$ & $ 0.21$ \\ 
      A496 & $209.59$ & $-36.49$ & $0.0328$ & $ 3.99$ & $1.49$ & $4.63$ & $ 0.14$ \\ 
    HydraA & $242.93$ & $ 25.09$ & $0.0549$ & $ 3.81$ & $1.45$ & $3.23$ & $ 0.07$ \\ 
     A4059 & $356.83$ & $-76.06$ & $0.0491$ & $ 3.79$ & $1.45$ & $2.39$ & $ 0.11$ \\ 
     A2199 & $ 62.90$ & $ 43.70$ & $0.0304$ & $ 3.73$ & $1.46$ & $2.99$ & $ 0.24$ \\ 
      A133 & $149.76$ & $-84.23$ & $0.0569$ & $ 3.45$ & $1.40$ & $1.74$ & $ 0.11$ \\ 
      A576 & $161.42$ & $ 26.24$ & $0.0401$ & $ 3.14$ & $1.37$ & $2.81$ & $ 0.12$ \\ 
    2A0335 & $176.25$ & $-35.08$ & $0.0346$ & $ 3.12$ & $1.37$ & $2.56$ & $ 0.18$ \\ 
     A2657 & $ 96.65$ & $-50.30$ & $0.0402$ & $ 3.00$ & $1.35$ & $2.54$ & $ 0.13$ \\ 
     A2063 & $ 12.85$ & $ 49.71$ & $0.0342$ & $ 2.96$ & $1.35$ & $1.85$ & $ 0.25$ \\ 
     A1736 & $312.58$ & $ 35.10$ & $0.0449$ & $ 2.81$ & $1.32$ & $3.14$ & $ 0.08$ \\ 
     MKW3s & $ 11.39$ & $ 49.45$ & $0.0453$ & $ 2.79$ & $1.31$ & $2.08$ & $ 0.12$ \\ 
     A2589 & $ 94.66$ & $-41.20$ & $0.0411$ & $ 2.59$ & $1.28$ & $1.30$ & $ 0.21$ \\ 
     A2052 & $  9.39$ & $ 50.10$ & $0.0345$ & $ 2.45$ & $1.27$ & $1.95$ & $ 0.19$ \\ 
     A2634 & $103.45$ & $-33.06$ & $0.0305$ & $ 2.31$ & $1.24$ & $2.62$ & $ 0.18$ \\ 
     A4038 & $ 24.89$ & $-75.82$ & $0.0288$ & $ 2.19$ & $1.22$ & $2.67$ & $ 0.19$ \\ 
   EXO0422 & $203.30$ & $-36.16$ & $0.0382$ & $ 2.00$ & $1.18$ & $4.67$ & $ 0.06$ \\ 
\hline
     A0754 & $239.25$ & $ 24.75$ & $0.0528$ & $13.57$ & $2.22$ & $4.43$ & $ 0.16$ \\ 
     A1367 & $234.80$ & $ 73.03$ & $0.0216$ & $ 7.08$ & $1.82$ & $4.61$ & $ 0.53$ \\ 
 Centaurus & $302.41$ & $ 21.56$ & $0.0499$ & $ 3.19$ & $1.37$ & $2.93$ & $ 0.08$ \\ 
      Coma & $ 58.09$ & $ 87.96$ & $0.0232$ & $ 9.60$ & $2.01$ & $2.79$ & $ 1.00$ \\ 
    Fornax & $236.72$ & $-53.64$ & $0.0046$ & $ 1.17$ & $1.01$ & $0.54$ & $21.51$ \\ 
     Hydra & $269.63$ & $ 26.51$ & $0.0114$ & $ 2.31$ & $1.26$ & $3.77$ & $ 0.89$ \\ 
       M49 & $286.92$ & $ 70.17$ & $0.0044$ & $ 0.60$ & $0.80$ & $0.93$ & $ 7.54$ \\ 
   NGC4636 & $297.75$ & $ 65.47$ & $0.0037$ & $ 0.16$ & $0.51$ & $1.25$ & $ 2.48$ \\ 
   NGC5044 & $311.23$ & $ 46.10$ & $0.0090$ & $ 0.43$ & $0.72$ & $1.54$ & $ 0.82$ \\ 
   NGC5813 & $359.18$ & $ 49.85$ & $0.0064$ & $ 0.38$ & $0.69$ & $3.06$ & $ 0.73$ \\ 
   NGC5846 & $  0.43$ & $ 48.80$ & $0.0061$ & $ 0.15$ & $0.51$ & $3.71$ & $ 0.31$ \\ 
\hline \hline
\end{tabular}
\end{center}
\end{table}

We choose 23 galaxy clusters that are the brightest in X-rays 
from Ref.~\cite{Reiprich2002}. These are the same clusters as those analyzed 
in Ref.~\cite{FermiCluster} for general gamma-ray analyses of galaxy clusters, 
where the sample clusters are chosen such that they yield the largest mass divided 
by distance squared, which is a reliable measure of expected annihilation fluxes.
In addition, we also analyze 34 massive low-redshift clusters from
the Chandra Cluster Cosmology Project~\cite{Vikhlinin2009a}.
We include this sample in order to investigate the level of cosmic-ray pressure 
in clusters and their implications for cluster-based cosmological tests~\cite{AndoNagai}, which we will discuss in more detail in our subsequent work.
We remove clusters located at low Galactic latitudes, $|b| < 20^\circ$, in 
the analysis because of the strong 
emission from the Galactic plane. Consequently, we have 49 clusters to 
analyze. Two of them (A2199 and A3571) are found in both samples.
Table~\ref{table:clusters} summarizes properties of these clusters.

In the table we also show the flux upper limits for each cluster assuming $E^{-2}$ 
spectrum and that the clusters are point-like. The seventh column shows the integrated 
flux upper limits in 0.1--100\,GeV, $F_{\rm lim}(>0.1 \, \mathrm{GeV})$, and they are 
typically $(1$--$5)\times 10^{-9} \, \mathrm{cm^{-2} \, s^{-1}}$.
The last column of the table shows the ratio of expected gamma-ray flux
from dark matter annihilation in the case of smooth NFW halo model to
the flux upper limits, with respect to the value for the Coma cluster.
These ratios, therefore, indicate the promise of each cluster as a target 
for dark matter annihilation.  It is clear from this table that the Fornax cluster 
is by far the most promising target among 49 clusters considered in this work.

\begin{figure}
\begin{center}
\includegraphics[width=8.0cm]{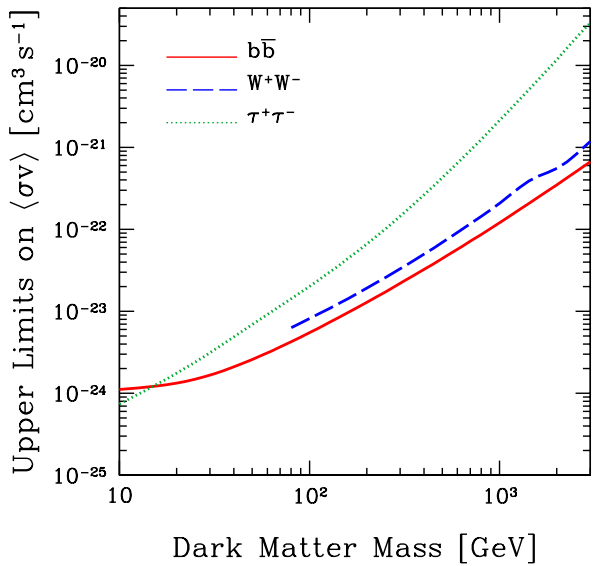}
\caption{The cross section upper limits from Fornax cluster (smooth NFW
 with no contraction)
 for the $b\bar b$, $W^+ W^-$, and $\tau^+ \tau^-$ annihilation
 channels.}
\label{fig:limit_channel}
\end{center}
\end{figure}

In Fig.~\ref{fig:limit_channel}, we show cross section upper limits as a
function of WIMP mass for the three different annihilation channels obtained
from observations of the Fornax cluster.
The channel into gauge bosons will give similar yield to $b\bar b$ for
all the masses, but limits for $\tau$-lepton pairs are significantly
weaker especially at larger masses, because of hard spectrum
(Fig.~\ref{fig:DM_spectrum}). We note that these limits are consistent with 
those of Refs.~\cite{FermiClusterDM, Huang2011} after correcting for 
differences in exposures.

\subsection{Stacking analysis}
\label{sec:stack}

One may wonder if stacking a large statistical sample of galaxy clusters 
improves upper limits on the annihilation cross section from those obtained 
individually.
%We shall show that with such an approach, one can improve the cross
%section upper limits by only $\sim$10--20\% for low-mass WIMPs.
%For the stacking analysis in this section, we simply adopt the NFW
%profiles (Sec.~\ref{sub:DM}).
Here, we stack a subset of 49 galaxy clusters discussed above, which is a 
significant improvement from the previous studies based on 6 or 
8 clusters~\cite{FermiClusterDM, Huang2011}.
The procedure for obtaining the stacking constraint on the cross 
section is similar to the individual cluster analysis (see Appendix~\ref{sec:analysis} 
for more details). In order not to bias the limits too high, we remove
clusters with signal-to-noise ratio larger than $3\sigma$ compared with the (fixed) background model for \textit{any} 
dark matter masses and annihilation channels.\footnote{Note that this does not 
mean that some clusters are already detected with $\gtrsim 3\sigma$, because 
there are uncertainties in our procedure of fixing background.}
%Although such a procedure is sufficient for the current
%purpose, for claiming detection, a more careful treatment should be made
%by varying parameters for the background models together with the
%cluster emission itself.}
As the result, we have 38 clusters used in the stacking analysis.

%  In general, the limits 
%improve if we gather statistical power by stacking an optimized number 
%of galaxy clusters.
%we optimize sub-sample of clusters to be used in the stacking
%analysis such that the obtained upper limits are minimized.
%For example, in the case of the NFW halo model and $b\bar b$ annihilation 
%channel, the optimized number of clusters used in the analysis is two (i.e.,
%Fornax and M49) when the mass of dark matter is less than 100~GeV and
%increases to six when the mass is more than 1~TeV.
%If we included more clusters than these, obtained upper limits would
%be weaker.
%Hence, although we included 49 clusters in our sample, most of them do
%not contribute to the stacking constraints on the annihilation cross section. 

\begin{figure}
\begin{center}
\includegraphics[height=15cm]{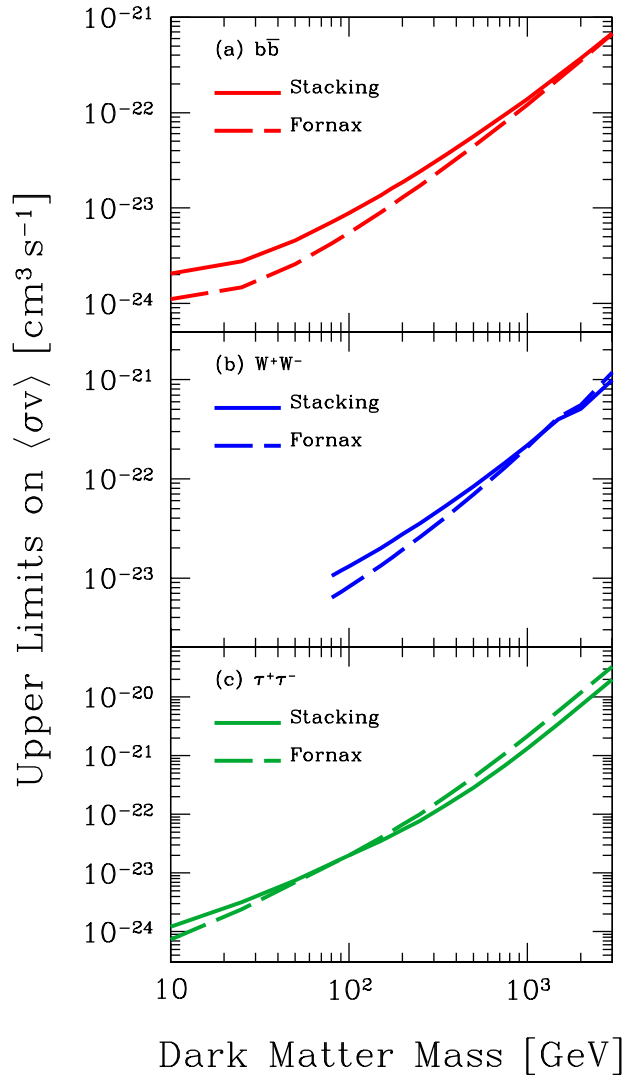}
\caption{Upper limits on annihilation cross section $\langle \sigma v
 \rangle$ for the $b\bar b$, $W^+W^-$, and $\tau^+\tau^-$ annihilation
 channels, in the case of smooth NFW host halo model (with no
 contraction). Solid and dashed curves are the limits due to staked
 clusters and Fornax, respectively.}
\label{fig:limit_stack}
\end{center}
\end{figure}

In Fig.~\ref{fig:limit_stack}, we compare the cross section upper
limits from the stacking analysis and the constraint from the Fornax
cluster alone for the three different annihilation channels.
We find that the stacked limits of the cross section degrade toward lower dark matter 
mass, while they improve by about a factor of 2 at multi-TeV mass regime 
for the $\tau^+ \tau^-$ channel.
This is in part because we still have weak (less than $3\sigma$) excess 
emission from some galaxy clusters for some annihilation models.
This shows that the Fornax cluster dominates the constraint on the 
annihilation cross section, and stacking a large sample of clusters does 
not help improve the limits.
This is in contrast to the case of dwarf satellite galaxies,
where the stacking analysis of 10 satellite galaxies can improve 
the limits by a factor of a few for a wide range of dark matter 
masses~\cite{FermiDwarfDM2, FermiDwarfDM3}.
In the next section, we shall discuss that treating the Fornax cluster
carefully is indeed more important for setting an improved constraint
on the annihilation cross section based on galaxy clusters. 
%, as the improvement by the stacking 
%analysis of clusters is much less than the improvement that we obtain 
%by carefully modeling the effect of dark matter contraction in Fornax.

%\section{Application to the Fornax cluster}
\section{Constraints from the Fornax cluster}
\label{sec:Fornax}

As shown in Table~\ref{table:clusters}, the Fornax cluster is located at
$(l, b) = (236.72^\circ, -53.64^\circ)$ in the Galactic coordinate. Its
redshift is $z = 0.0046$, and the corresponding angular diameter distance 
is $d_A = 13.7\, h^{-1}\, \mathrm{Mpc}$.  Both optical and X-ray observations show that 
the brightness profiles of Fornax are fairly regular, and therefore, modeling 
dark matter distribution assuming spherical symmetry is well justified.

\subsection{Distribution of dark matter, stars, and gas}

The X-ray brightness has been used to infer the mass to be $M_{500} =
6.4 \times 10^{13} \, h^{-1}\, M_\odot$~\cite{Chen2007, Reiprich2002}
(see also Refs.~\cite{Churazov2008, Das2010}), and the corresponding
virial mass and concentration are $M_{\rm vir} = 1.2 \times 10^{14}\,
h^{-1}\, M_\odot$ and $c_{\rm vir} = 5.6$, respectively.
For the Einasto profile, Eq.~(\ref{eq:Einasto}), we adopt $n = 3.87$ and
$r_e = 1247\, \mathrm{kpc}$, following the cluster simulation of similar
size (model C09 of Ref.~\cite{Merritt2006}).

We shall obtain $M_\star(r)$ and $M_{\rm gas}(r)$ as they are necessary
ingredients for calculating the final dark matter profile after contraction 
using Eq.~(\ref{eq:contra}).  For the stellar mass distribution, 
we use the HST observation of the bright elliptical galaxy, NGC~1399, 
located at the center of the Fornax cluster.
The surface brightness profile of stars are well represented by the 
``Nuker law''~\cite{Lauer1995}:
\begin{equation}
 I(\theta) = 2^{(\beta - \gamma) / \alpha} I_b
  \left(\frac{\theta}{\theta_b}\right)^{-\gamma}
  \left[1 + \left(\frac{\theta}{\theta_b}\right)^\alpha\right]
  ^{(\gamma-\beta)/\alpha},
  \label{eq:Nuker}
\end{equation}
which behaves as $\theta^{-\gamma}$ ($\theta^{-\beta}$) at small (large)
$\theta$ with a break around $\theta_b$.
We here consider the results of HST WFPC2 observations using the F606W
filter~\cite{Lauer2005, Lauer2007}: $\alpha = 1.58$, $\beta = 1.63$,
$\gamma = 0.09$, $\theta_b = 3.17^{\prime\prime}$, and $I_b =
17.23\, \mathrm{mag\, arcsec^{-2}}$.
This photometry result is valid down to a very small radii, $\sim$1\,pc
(i.e., $\sim$0.01$^{\prime\prime}$).
The same measurements also constrain the ellipticity of the galaxy,
which is found very small~\cite{Lauer2005}.  Given that the assumption
of spherical symmetry is well justified, we obtain the luminosity density 
by de-projection~\cite{Binney2008}:
\begin{equation}
 j(r) = -4 \int_{r/d_A}^{\infty}
  \frac{I^\prime(\theta)d\theta}{\sqrt{\theta^2 d_A^2 - r^2}}.
\end{equation}
In order to obtain the density profile, we finally multiply $j(r)$ by
the mass-to-light ratio, $\rho_\star(r) = (M/L) j(r)$.
According to the stellar kinematics study, it is believed that the
gravitational potential in the central regions is dominated by stellar
component, and the mass-to-light ratio has been inferred to be about 10
times that of the Sun~\cite{Saglia2000, Kronawitter2000}, which is
typical for old stellar population. 
Figure~\ref{fig:profile} shows the density profile of the stellar
component, $\rho_\star$, obtained in this way, and it dominates the mass
within $\sim$10\,kpc.
We note that the power-law extrapolation of Eq.~(\ref{eq:Nuker}) works
reasonably well even at large radii, but underestimates the brightness
(and therefore gives conservative results) beyond
$\sim$20\,kpc~\cite{Killeen1988}.

The thermal gas component can be inferred from X-ray data.
Here we use the ROSAT observations of the Fornax
cluster~\cite{Jones1997, Paolillo2002}.
In particular, Ref.~\cite{Paolillo2002} identified three different
components and fitted each with a bi-dimensional $\beta$-profile.
%Although there are offsets between the central positions of these
%different components, they are much smaller than the point spread
%functions of the gamma-ray telescopes, and therefore can safely be
%ignored.
The thermal gas density and mass profile are shown 
in Figure~\ref{fig:profile}. They are subdominant compared to 
the stellar density and mass within $\sim$100\,kpc, and hence 
do not play important role in the dark matter contraction directly.
However, since it is the dominant baryon component around the 
virial radius, it affects the dark matter contraction calculation 
through $f_b$ in an indirect way.
%~\cite{Gnedin2004, Gnedin2011}.

%\subsection{Results for Fornax}

\subsection{Effects of dark matter contraction}

%\subsection{Dark matter density profile after contraction}

\begin{figure}
\begin{center}
\includegraphics[width=7cm]{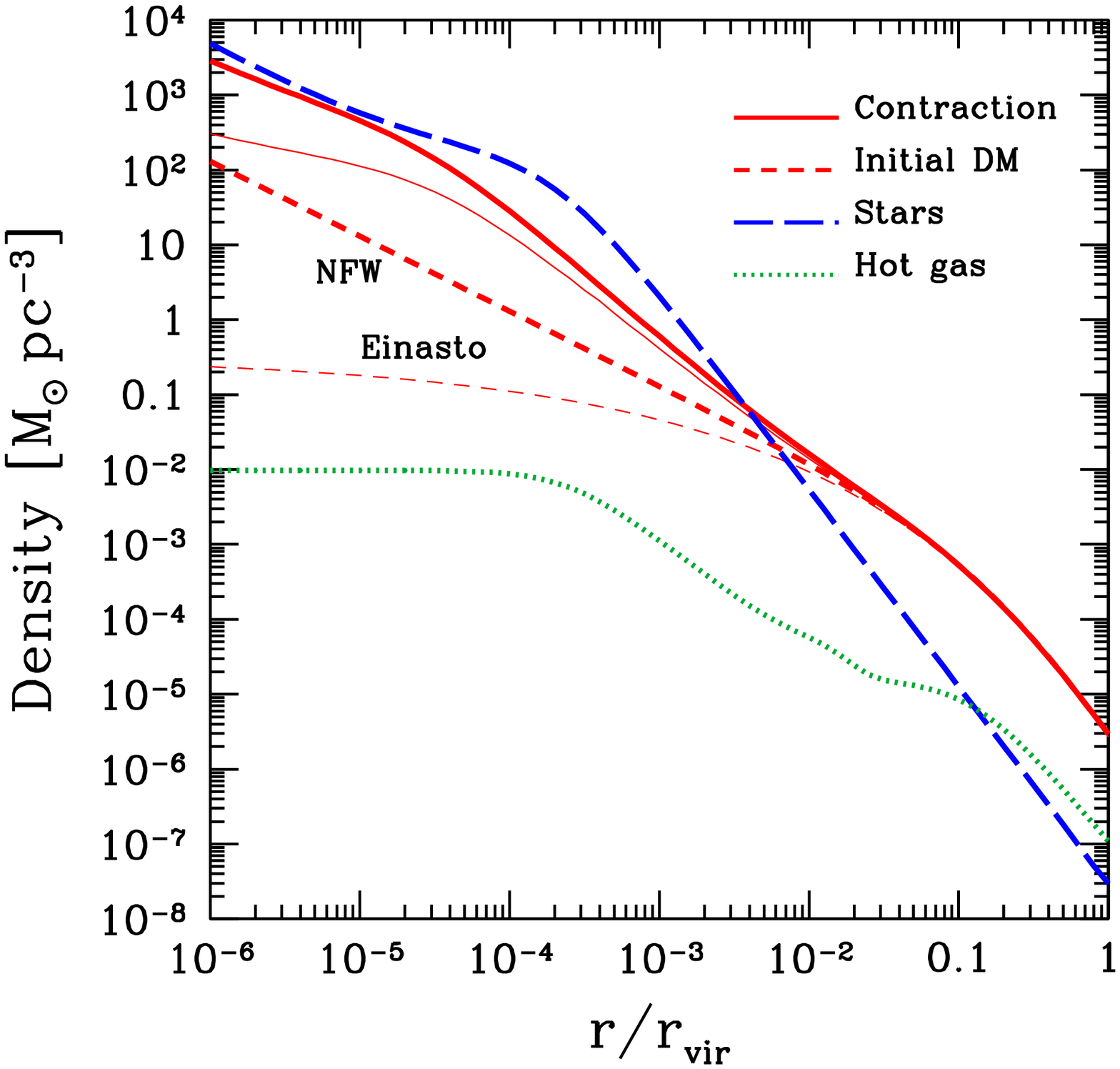}
\hspace{0.5cm}
\includegraphics[width=7cm]{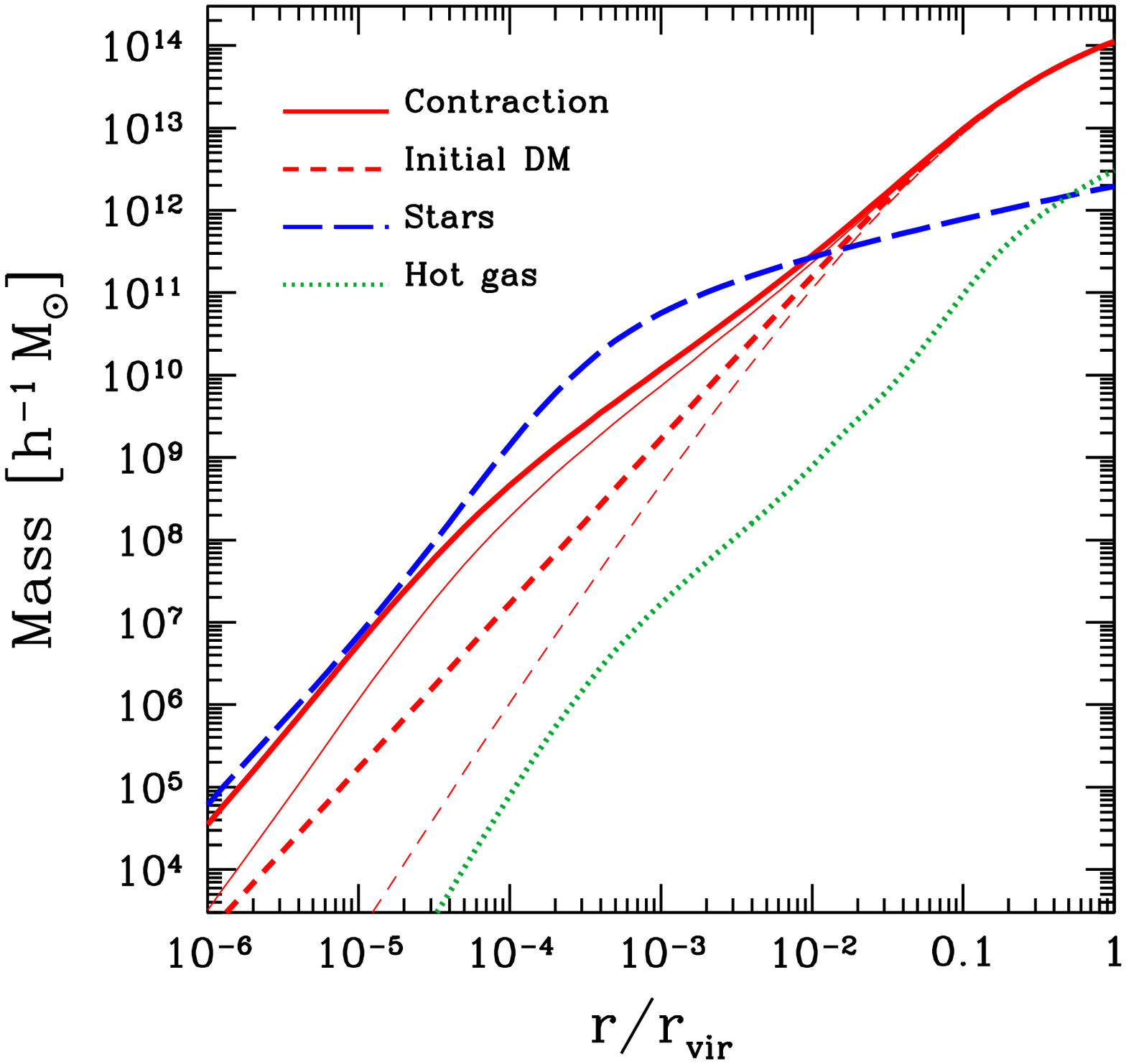}
\vspace{-0.5cm}
\caption{Density (left) and mass (right) profiles of dark matter (before
 contraction in short-dashed line; after contraction in solid line), stars
 (long-dashed), and baryonic gas (dotted). For dark matter before and after
 contraction, both the NFW (thick) and Einasto (thin) profiles are
 shown. The canonical parameters ($A_0 = 1.6$ and $w = 0.8$) are adopted
 for the contraction model.}
\label{fig:profile}
\end{center}
\end{figure}

In Fig.~\ref{fig:profile}, we show both initial and final dark matter
density profiles for the Fornax cluster.
We consider the NFW and Einasto models as the initial dark
matter profile for the Fornax cluster and adopt the canonical
contraction model ($A_0 = 1.6$ and $w = 0.8$).
The initial dark matter profiles are in good agreement with each other at 
radii larger than $\sim$10~kpc, beyond which the total mass has been measured. 
Although the dark matter density is still subdominant at the smaller radii
compared to the baryon density dominated by stars, it is enhanced
significantly because of the dark matter contraction due to baryonic infall. 
This is the case even if the original profile has an inner core as 
the Einasto profile.
%We assume that this result does not affect the
%conclusions of the earlier studies on mass-to-light ratio of NGC~1399 
%such as Ref.~\cite{Saglia2000}, where dark matter was argued to have a flat
%core at the central regions. 
%This assumption will perhaps remain valid even in the cases with
%contractions, because stars are still the most dominant component to the
%gravitational potential; even in the case of the initial NFW profile,
%dark matter contribution is still less than $\sim$40\% of the total mass
%within 10\,kpc, where Ref.~\cite{Saglia2000} concluded that the stars
%dominate the potential.
%However, we note that this problem is worth revisiting, although it is
%beyond the scope of the present paper. 

\begin{figure}
\begin{center}
\includegraphics[width=8cm]{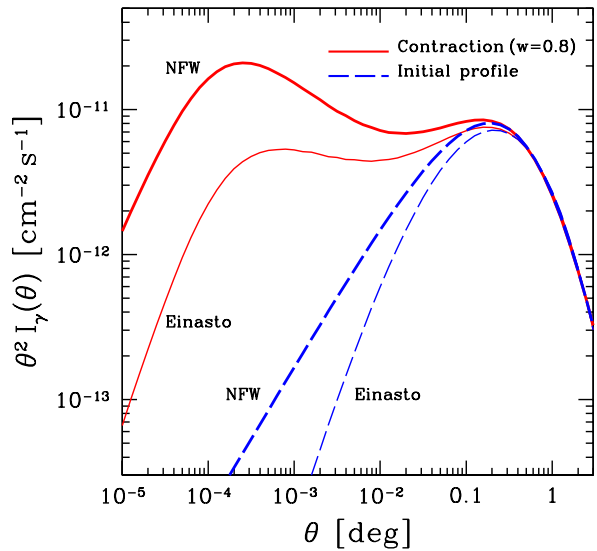}
\vspace{-0.5cm}
\caption{Gamma-ray intensity as a function of angle from the cluster
 center before and after the dark matter contraction, for both the NFW
 and Einasto profiles. Note that the instrumental responses have not
 been included yet. The particle-physics parameters are fixed to be
 $N_{\gamma, {\rm ann}}\langle \sigma v\rangle/m_\chi^2 = 10^{-26}\,
 \mathrm{cm^3 \, s^{-1} \, GeV^{-2}}$.}
\label{fig:intensity_f}
\end{center}
\end{figure}

Once the density profile is obtained, we compute the gamma-ray intensity 
using Eq.~(\ref{eq:intensity}).
Figure~\ref{fig:intensity_f} shows the intensity integrated over energy,
$I_\gamma(\theta)$, for both the NFW and Einasto profiles as initial
density models.
Here, we fix particle-physics parameters to be $N_{\gamma, {\rm ann}}
\langle \sigma v\rangle /m_\chi^2 = 10^{-26}\, \mathrm{cm^3 \, s^{-1} \,
GeV^{-2}}$.
In order to see what range of $\theta$ contributes to the total signals
the most, we multiply $I_\gamma(\theta)$ by $\theta^2$, such that the
area below each curve gives the gamma-ray flux.
One notices that there are double-peak structures.
One peak at higher $\theta$ corresponds to the scale
radius $r_s$ of the NFW or Einasto profiles. The other peak, on the 
other hand, corresponds to the characteristic scale of the dark 
matter contraction. 
Since the typical angular resolutions of the current generation of
gamma-ray telescopes are on the order of 0.1--1$^\circ$ (depending on
energies of gamma rays), the peaks at lower $\theta$ will simply make a
central pixel brighter but do not make additional resolvable features in
the maps.

\begin{figure}
\begin{center}
\includegraphics[width=7cm]{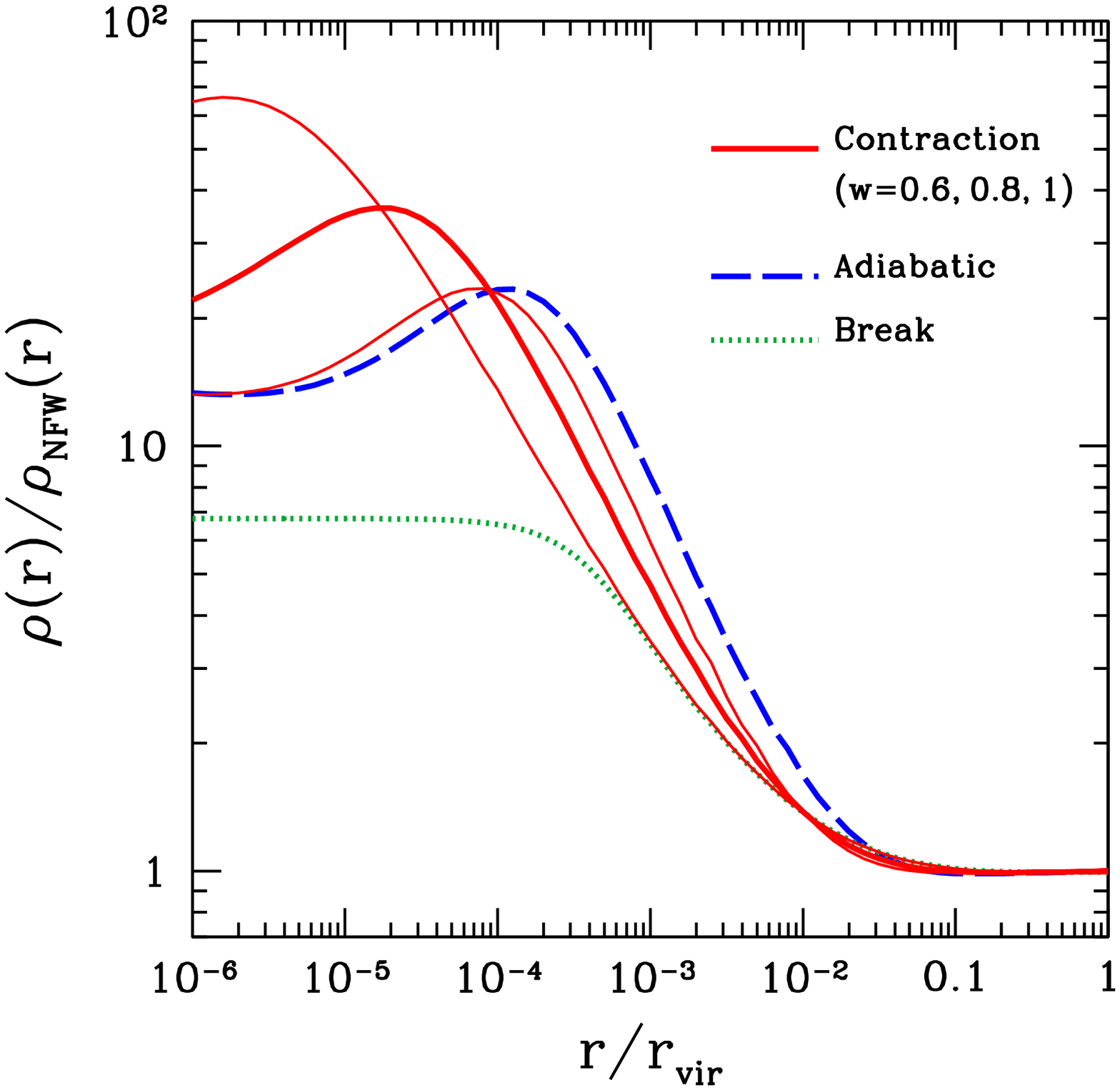}
\hspace{0.5cm}
\includegraphics[width=7cm]{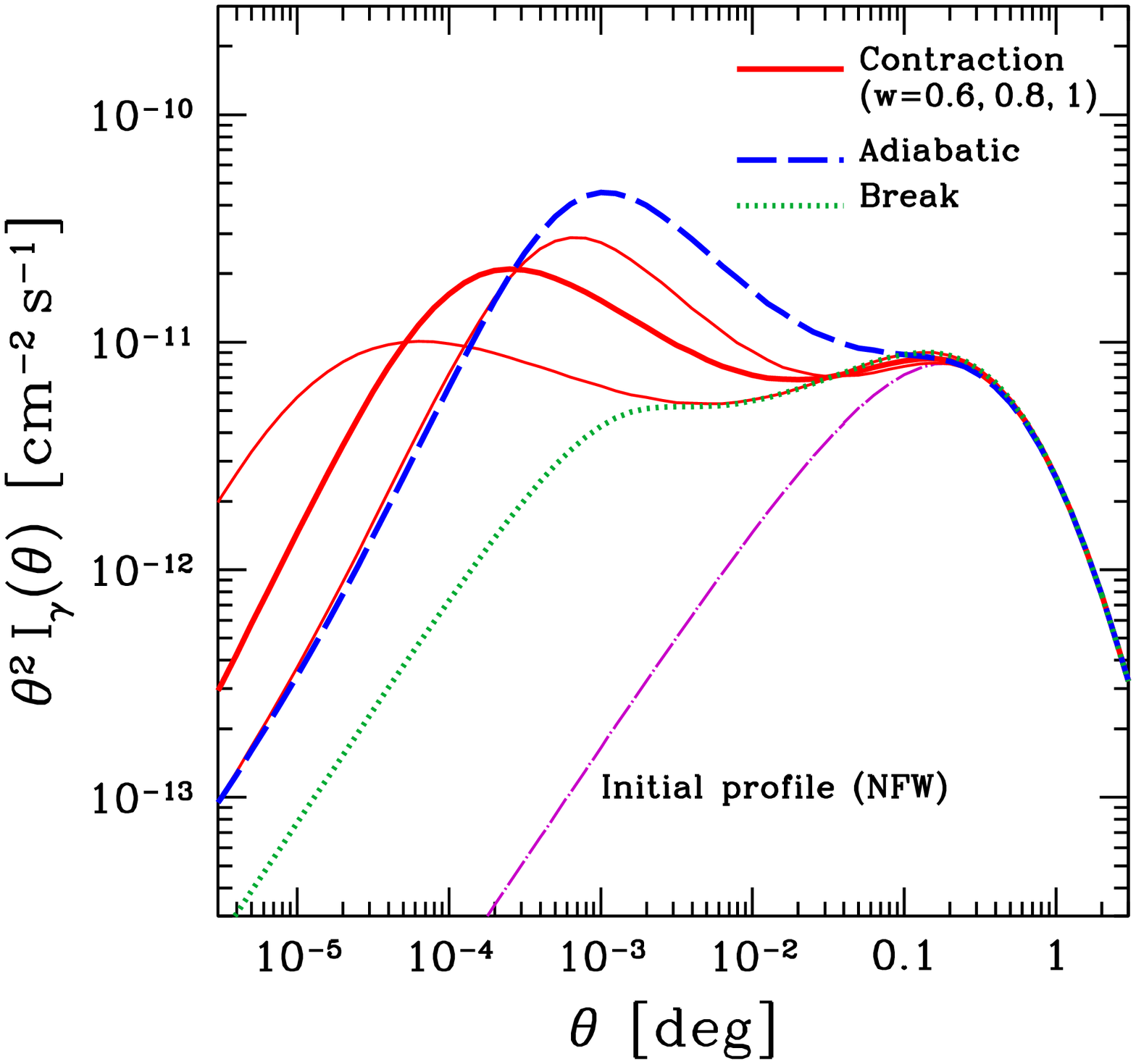}
\vspace{-0.5cm}
\caption{Density enhancement of dark matter, $\rho_\chi(r) / \rho_{\rm
 NFW}(r)$ (left) and gamma-ray intensity as a function of angle 
 from the cluster center (right), due to the contraction for the initial 
 NFW model.  Solid curves
 are for the power-law models of $\bar r$ [Eq.~(\ref{eq:rbar})], with $w
 = 0.6$ (top at low radii), 0.8 (middle), and 1 (bottom). Dashed line is
 for the standard adiabatic contraction model ($A_0 = 1, w = 1$), and
 dotted line is for $A_0 = 1.6, w = 0.6$ but with a power-law break in
 Eq.~(\ref{eq:rbar}) such that the $\bar r$ gets constant at $10^{-2}
 r_{\rm vir}$ for small radii.}
\label{fig:density_f_contra}
\end{center}
\end{figure}

The left panel of Fig.~\ref{fig:density_f_contra} shows a density 
enhancement factor after dark matter contraction, 
$\rho_{{\rm dm}, f} / \rho_{{\rm dm},i}$, for various dark matter contraction 
models. Here we assume the NFW model as an initial density profile. 
We compare results for three different values of $w = 0.6$, 0.8, and 1
by fixing $A_0 = 1.6$ in Eq.~(\ref{eq:rbar}).
The larger the parameter $w$ is, the more efficient the contraction
gets, since $\bar r$ is smaller. We also show results of two extreme scenarios.
One is the standard adiabatic contraction model ($A_0 = 1$ and $w = 1$),
and the other is the broken power-law model where $\bar r$ follows
Eq.~(\ref{eq:rbar}) with $w = 0.6$ at large radii but becomes constant
($\bar r = 10^{-2} r_{\rm vir}$) for small radii [see discussions after
Eq.~(\ref{eq:rbar})]; they are labeled as ``Adiabatic'' and ``Break,''
respectively. The true model is likely bracketed by these extreme cases.

The right panel of Fig.~\ref{fig:density_f_contra} shows the gamma-ray 
intensities for the various contraction models considered above. 
As shown in the left panel of Fig.~\ref{fig:density_f_contra}, the flux
enhancements are bracketed by the two extreme models.
Although the peak moves towards larger $\theta$ for smaller $\bar r$,
one cannot resolve such peak yet with the current gamma-ray telescopes.
The inner peaks are around $10^{-5}$--$10^{-3}\,{\rm deg}$, which
corresponds to 3--300\,pc in the physical size. By integrating over $\theta$ 
out to large values, we find that the gamma-ray flux is boosted by a factor 
of 2--7 by the dark matter contraction effect.

%In Fig.~\ref{fig:cluster maps}, we show what the Fornax cluster
%map would appear when it is just as bright as currently allowed and is
%dominated by smooth NFW component.
%This is obtained by the likelihood analysis with the data shown in
%Fig.~\ref{fig:counts_map_Fornax}.
%The photon counts per $0.02^\circ \times 0.02^\circ$ pixel could be at
%most 0.017 at the center of the map.

%In Fig.~\ref{fig:limits_channel}, we compare upper limits for the three
%different annihilation channels, in the case of smooth Fornax model.
%The channel into gauge bosons will give similar yield to $b\bar b$ for
%all the masses, but limits for $\tau$-lepton pairs are significantly
%weaker especially at larger masses, because of very hard spectrum
%(Fig.~\ref{fig:DM_spectrum})

%\subsection{Constraints on dark matter annihilation cross section}
%\label{sub:results}

\begin{figure}
\begin{center}
\includegraphics[height=15cm]{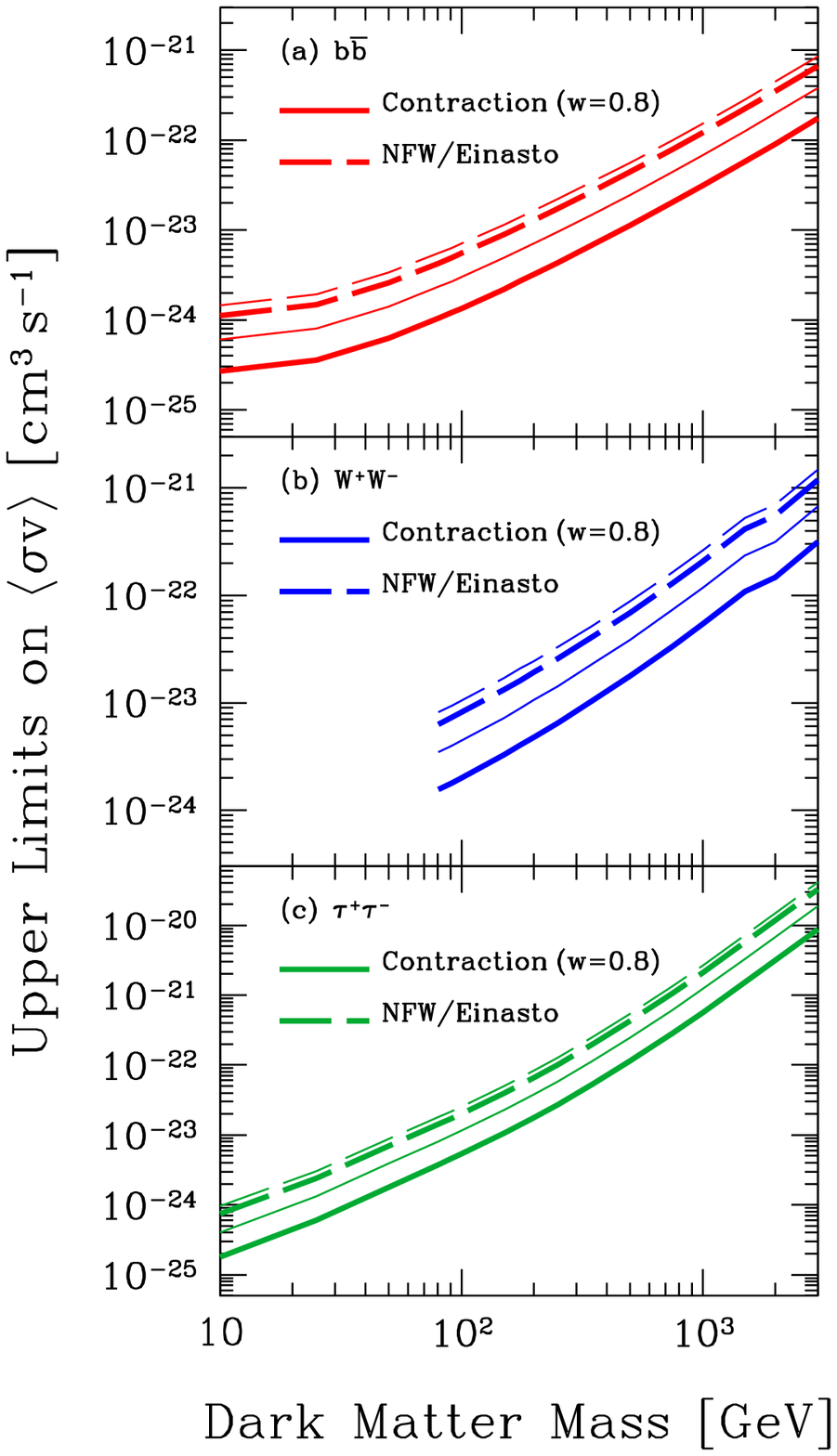}
\caption{Upper limits (corresponding to 95\% credible interval) on annihilation cross section $\langle
 \sigma v \rangle$ as a function of dark matter mass. The limits for the
 initial profiles (no-contraction model) are compared with those for the
 canonical contraction model with $A_0 = 1.6$ and $w = 0.8$. Thick and
 thin curves correspond to the NFW and Einasto profiles, respectively.}
\label{fig:limit_Fornax}
\end{center}
\end{figure}

We show, in Fig.~\ref{fig:limit_Fornax}, the upper limits (corresponding
to 95\% credible interval) on the
annihilation cross section, $\langle \sigma v \rangle$, as a function of
dark matter mass for the initial NFW and Einasto profiles, and for
various annihilation channels.
We compare no contraction model with the canonical contraction model
($A_0 = 1.6$ and $w = 0.8$).
The contraction model improves the limits at $m_\chi = 100\,{\rm GeV}$
by a factor of 4.1 for the $b\bar b$ and $W^+ W^-$ channels, and 3.8 for
the $\tau^+ \tau^-$ channel in the case of the NFW initial profile, and
this is almost independent of mass.
In the case of the Einasto initial profile, on the other hand, the
improvement factors at 100\,GeV are 2.4 for $b\bar b$ and $W^+ W^-$, and
2.2 for $\tau^+ \tau^-$.
We note that the limits for no-contraction models of the NFW and Einasto
profiles differ by only $\sim$20\% (the former is better), which is
negligible compared to the differences between models with and without 
contraction.

\begin{figure}
\begin{center}
\includegraphics[width=8.5cm]{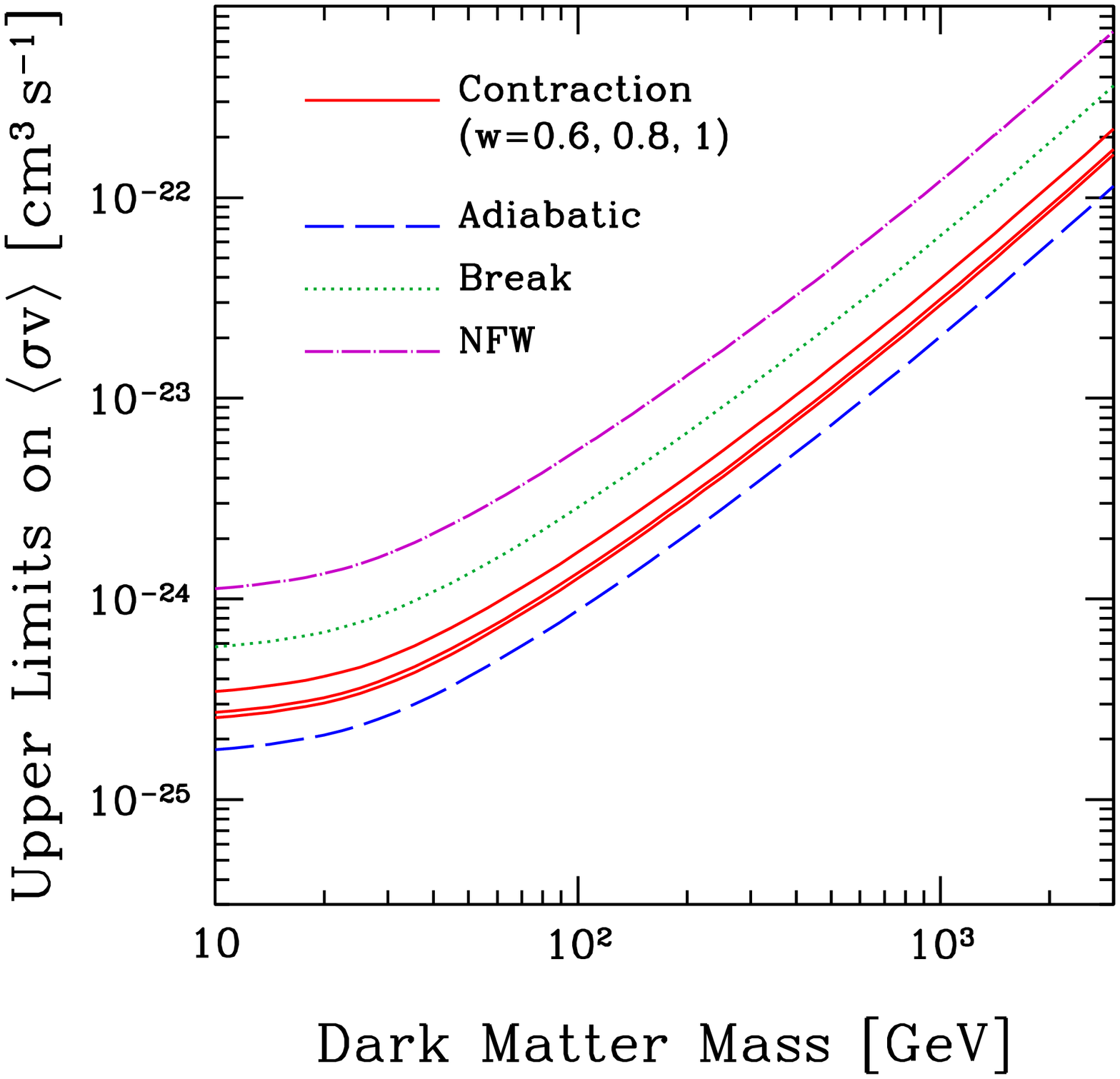}
\caption{Cross section upper limits $\langle \sigma v \rangle$ for
 various contraction models for the NFW initial profile and the $b\bar
 b$ annihilation channel. The line types are the same as those in
 Fig.~\ref{fig:density_f_contra}.}
\label{fig:limit_contra}
\end{center}
\end{figure}

We compare various contraction models for the NFW initial profile and
the $b\bar b$ annihilation channel in Fig.~\ref{fig:limit_contra}.
As expected, the improvement factor is bracketed by our two extreme
models, between 1.9 and 6.3 for the mass of 100\,GeV and the $b\bar b$
channel.
It is worth pointing out that even the very conservative model of the
contraction with the power-law break at small radii in
Eq.~(\ref{eq:rbar}) gives a factor of 2 enhancement compared to the
no-contraction model, highlighting the importance of dark matter 
contraction in interpretation of the dark matter annihilation signals 
from galaxy clusters. This effect will make clusters brighter in a manner 
independent of the boost due to subhalos, which is the effect of much 
interest in the past several years and revisited below.

\subsection{Effects of substructure boost and minimum mass}

\begin{figure}
\begin{center}
\includegraphics[width=8cm]{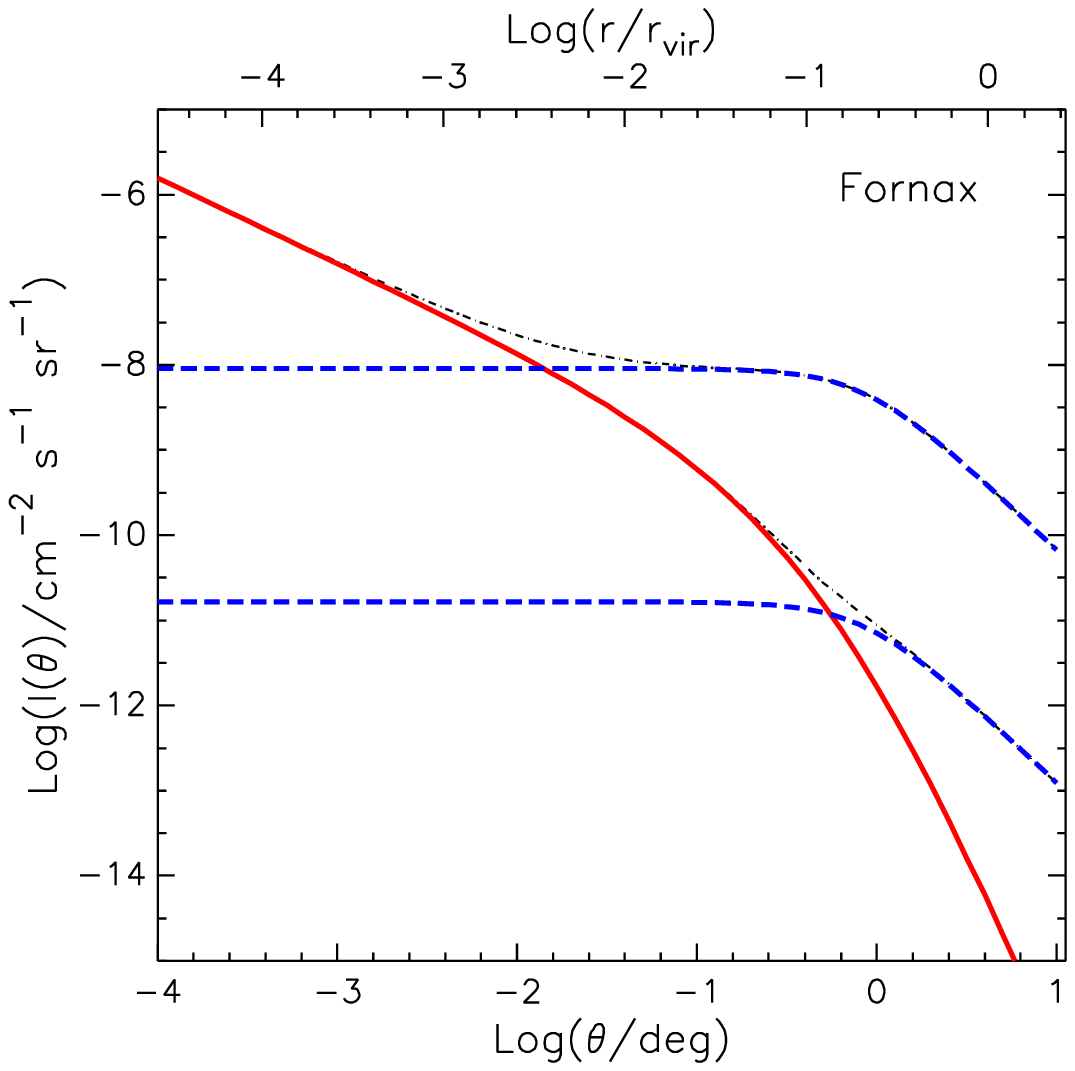}
\caption{Intensity $I(\theta)$ as a function of angle $\theta$ (and
 radius in units of $r_{\rm vir}$) for the Fornax cluster. Dark matter
 parameters are fixed to be $N_{\gamma, {\rm ann}} \langle \sigma v
 \rangle / m_\chi^2 = 3\times 10^{-30} \, \mathrm{cm^3 \,
 s^{-1}}$. Solid curve is the contribution from smooth NFW component,
 and dashed curves are those from subhalos assuming minimum subhalo mass
 of $M_{\rm min} = 5\times 10^7 M_\odot$ (lower) and $M_{\rm min} =
 10^{-6} M_\odot$ (upper). Total intensities are shown as dot-dashed
 curves.}
\label{fig:intensity}
\end{center}
\end{figure}

If the gamma rays are dominated by annihilation in subhalos and the
subhalo mass function extends down to Earth mass (i.e., $M_{\rm min} =
10^{-6} M_\odot$), then the cross section upper limits are significantly 
improved.
Figure~\ref{fig:intensity} shows the intensity $I_\gamma(\theta)$
integrated over energy, as a function of angle $\theta$, for the Fornax
cluster.
Here, we fix the particle-physics parameter to be $N_{\gamma, {\rm ann}}
\langle \sigma v \rangle / m_\chi^2 = 3\times 10^{-30} \, \mathrm{cm^3
\, s^{-1} \, GeV^{-2}}$.
We show the contributions from both the smooth NFW host halo and
substructures.
For the latter, we adopt two minimum subhalo masses: $5\times 10^{7}
M_\odot$ corresponding to the current resolution limit~\cite{Gao2011} and 
$10^{-6} M_\odot$ corresponding to the typical smallest subhalo mass for 
the neutralino dark matter.
One can see that the subhalo contribution could exceed that of the host
halo if the subhalo mass function extends down to the Earth-mass scale. 
However, the relative importance of the host vs. subhalo contribution 
is highly sensitive to the cutoff mass scale. For example, if it is around the 
current resolution limit of simulations, then the host halo contribution 
becomes more dominant. 

\begin{figure}
\begin{center}
\includegraphics[width=8.0cm]{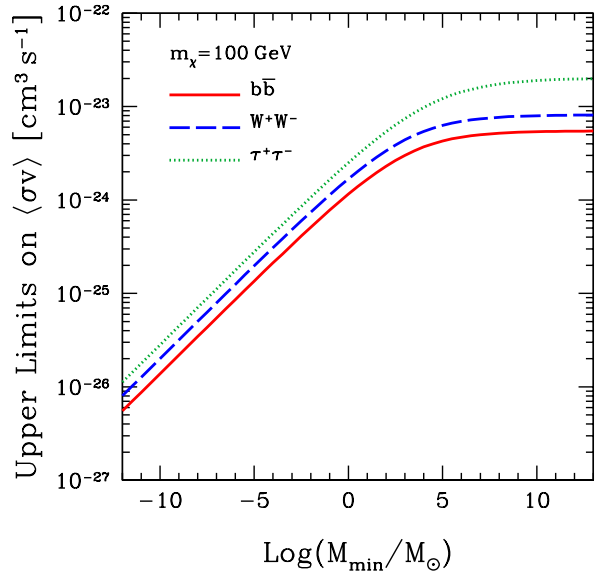}
\caption{Upper limits on annihilation cross section $\langle \sigma v
 \rangle$, obtained with the Fornax cluster, as a function of minimum
 subhalo mass, $M_{\rm min}$, for the three different annihilation
 channels and $m_\chi = 100 \, {\rm GeV}$. For the smooth component, the
 NFW profile with no contraction is adopted.}
\label{fig:limit_Mmin}
\end{center}
\end{figure}

Next, we investigate the dependence of the upper limit of dark matter
annihilation cross section from Fornax on the assumed minimum subhalo 
mass $M_{\rm min}$.  Results are shown in Fig.~\ref{fig:limit_Mmin} for 
$m_\chi = 100$~GeV and the three different annihilation channels.  
%, adopting the $M_{\rm
%min}^{-0.2}$ dependence of the boost factor $b$.
Here we adopt the NFW profile with no contraction for the smooth component.
Assuming the $b \propto M_{\rm min}^{-0.2}$ dependence, we find that
the upper limits on $\langle \sigma v \rangle$ scales as $M_{\rm
min}^{0.2}$, when $M_{\rm min}$ is small and the emission is dominated by
annihilation in subhalos.
This is realized for $M_{\rm min} \lesssim 10^2 M_\odot$.
For larger $M_{\rm min}$ than this value, the emission is dominated by
the smooth NFW host-halo component and the limits asymptotically approach 
the values evaluated assuming no substructure.
The canonical cross section of $3 \times 10^{-26} \, \mathrm{cm^3 \,
s^{-1}}$ is excluded, only if $M_{\rm min}$ is smaller than the Earth
mass.
We once again stress that the estimate of substructure boost
factor (by Earth-size dark matter subhalos) requires extrapolation of the
subhalo mass function by more than 13 orders of magnitude, given that the 
state-of-the-art numerical simulations can resolve halos of only 
$M \sim 5\times 10^7 M_\odot$ level~\cite{Gao2011}.

Comparing the relative importance of substructure boost to the effect
of dark matter contraction, we conclude that the substructure boost could
exceed the contraction boost, {\it only if} the minimum subhalo mass is 
considerably smaller than one solar mass. Otherwise, the substructure 
boost is at most a factor of a few, and the effect is generally smaller than 
the effects of contraction that ranges between a factor of 2 and 6. While
the substructure boost remains highly uncertain, it is critical to take
into account the contraction by baryon dissipation when studying 
the dark matter annihilation signals from clusters. 
%The current resolution of the numerical simulations are unfortunately
%far larger than the solar mass scale, and this will keep being so at
%least for the near future.
%Therefore, one has to seriously take the contraction into account when
%galaxy clusters are discussed for dark matter annihilation.

\section{Conclusions}
\label{sec:conclusions}

We investigated dark matter annihilation in galaxy clusters and
constraints on annihilation cross sections with gamma-ray data 
from Fermi-LAT. Our main results are summarized below.

%\begin{itemize}

%\item 
By analyzing 49 nearby massive galaxy clusters located at high
Galactic latitudes, we show that the Fornax cluster provides the most 
stringent constraints on the dark matter
annihilation cross section. In addition, we show that the stacking
analysis does not help improve the cross section upper limits
much. Our results suggest that detailed mass modeling of the 
Fornax cluster is more important for improving the dark matter 
annihilation cross section constraints from clusters than improving
the cluster sample size. 
%We, therefore, conclude that rather than analyzing a large
%statistical sample of galaxy clusters, we should instead model 
%the Fornax cluster precisely to obtain better estimates of the
%annihilation signatures.

%\item 
We therefore performed a detailed mass modeling and predicted
      expected dark matter annihilation signals of the Fornax cluster, by
      taking into account effects of dark matter contraction and
      substructures.  While the latter has been considered in the
      literature, the effect of dark matter contraction by baryon
      dissipation has not been considered in interpretation of the dark
      matter annihilation signal in cluster-size halos to date. By modeling the mass
      distribution of baryons (stars and gas) around a central bright
      elliptical galaxy, NGC~1399, and using a modified contraction
      model motivated by numerical simulations, we show that 
      the dark matter contraction boosts
      annihilation signatures by a factor of $\sim$4. Within a
      conservative range of contraction parameters, we also show that
      the flux boost will fall between a factor of 2 and 7.

%\item 
We analyzed the Fermi-LAT data for 2.8 years around the Fornax
      cluster. After taking all the sources and diffuse backgrounds
      into account, we obtained upper limits on the annihilation cross
      section for various models of contraction. We showed that the
      limits are improved, compared to those for the NFW or Einasto
      profiles with no contraction, by a factor of 2--6 for a wide range
      of parameters of the contraction model. For the canonical
      contraction model, the current limits are $\langle \sigma v \rangle
      \lesssim (2$--$3) \times 10^{-25} \, \mathrm{cm^3 \, s^{-1}}$ for
      dark matter with mass around 10\,GeV, which is roughly one order
      of magnitude larger than the
      canonical cross section necessary to explain the thermal relic
      density. We argue that this effect is more robust than the
      subhalo boost that has been discussed in the literature, and
      indeed, more important unless the minimum subhalo mass is 
      considerably smaller than the mass of the Sun.

%\end{itemize}

\acknowledgments
We thank Eiichiro Komatsu for enlightening discussions, Roberto Trotta
for helpful discussions, and Savvas Koushiappas for comments. SA was 
supported by the GRAPPA initiative at University of Amsterdam 
and in part by Japan Society for Promotion of Science.
DN was supported in part by NSF grant AST-1009811, by NASA 
ATP grant NNX11AE07G, and by the Yale University.

\appendix

\section{Analysis of Fermi-LAT data}
\label{sec:analysis}

%Here we illustrate analysis details, while the analysis results are
%presented in Sec.~\ref{sec:Fornax} and \ref{sec:stack}.
%whoThose who are interested in key results may skip to Sec.~\ref{sec:results}. 

\subsection{Photon data processing}

We perform the analysis using the Fermi {\it Science Tools} (v9r23p1), as well 
as data of gamma-ray photons collected by
Fermi-LAT for the mission elapsed time (MET) between 239557417~s and
329159098~s (2.8~years; from August 4, 2008 to June 7, 2011), contained
in a region of interest (ROI), i.e., a circle of 10$^\circ$-radius
around a target cluster.
First, we chose {\tt DIFFUSE} and {\tt DATACLEAN} class photons whose
energies are between 1~GeV and 100~GeV by using {\it gtselect}, such
that contamination of charged particles gets reduced to minimum.
Then with {\it gtmktime} we further cut the data according to the
spacecraft configuration, using ``{\tt DATA\_QUAL==1 \&\& LAT\_CONFIG==1
\&\& ABS(ROCK\_ANGLE)<52}'' filter as well as ROI-based zenith angle cut.

\subsection{Maximum likelihood analysis}

\begin{figure}
\begin{center}
\includegraphics[height=6.5cm]{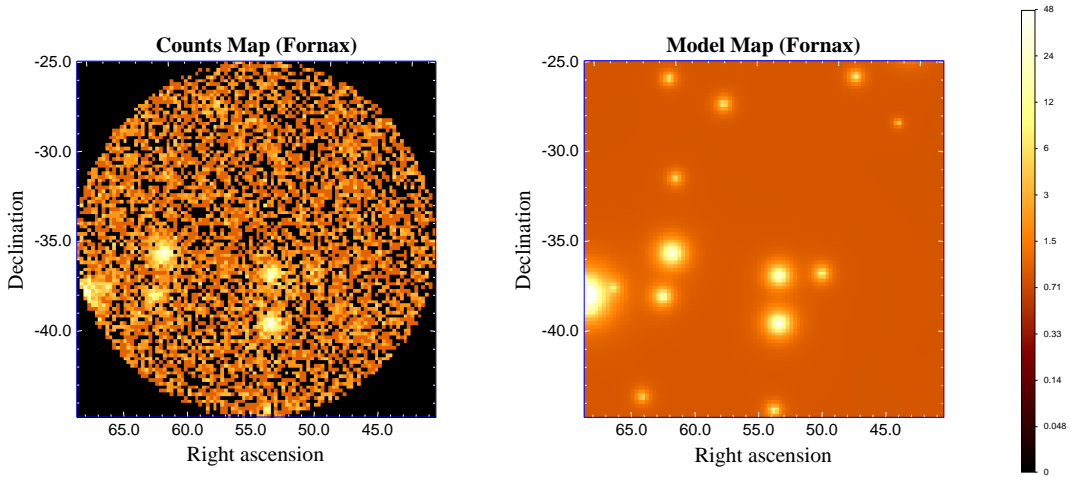}
\caption{Counts map (left) and model map (right) for 1--100~GeV photons
 around the Fornax cluster. The colors show source counts per pixel and
 are logarithmically spaced. Ten 2FGL point sources are found in ROI
 ($10^\circ$ radii) and 28 total are found in the source region
 ($15^\circ$ radii), in addition to the both Galactic and extragalactic
 diffuse backgrounds. Note that the flux of the source at far left in
 the model back is saturated.}
\label{fig:maps_Fornax}
\end{center}
\end{figure}

Using these filtered data around the cluster, we perform the binned
likelihood analysis, by comparing the photon data with the models of
sources (both point-like and extended), as well as the Galactic
foreground and extragalactic background.
The sources are taken from 2-year source catalog (2FGL)~\cite{2FGL}, if
they are located within 15$^\circ$ radii (source region) from the
cluster.
With {\it gtbin}, we divide the cluster map into spatial $100 \times
100$ pixels, whose size is $0.2^\circ \times 0.2^\circ$, and also into
20 energy bins equally spaced logarithmically between 1~GeV and
100~GeV.
The left panel of Fig.~\ref{fig:maps_Fornax} shows such a map (for
Fornax) of photon counts per pixel, but integrated over the entire
energy range, 1--100~GeV.
Based on this binning, with {\it gtltcube} followed by {\it gtexpcube2},
we produce exposure maps corresponding to each cluster, and then with
{\it gtsrcmaps}, maps of the sources and the diffuse backgrounds in
ROI.

We use {\it gtlike} to perform the maximum likelihood analysis, first
with {\tt DRMNFB} then followed by {\tt NEWMINUIT} optimizers.
This way, we constrain amplitudes of both the Galactic foreground and the
extragalactic background, and also for most of the sources, the flux and
the power-law index of the spectra if they are within ROI.
If the sources are out of ROI but still within source region, then the
fixed values from 2FGL catalog for these parameters are used.
Then finally, by combining results of likelihood analysis, we create the
model maps with {\it gtmodel}, to be compared with the photon count
maps.
The right panel of Fig.~\ref{fig:maps_Fornax} shows the best-fit model
map of the data shown in the left panel.

\subsection{Setting upper limits on cluster emission}

\begin{figure}
\begin{center}
\includegraphics[width=7.0cm]{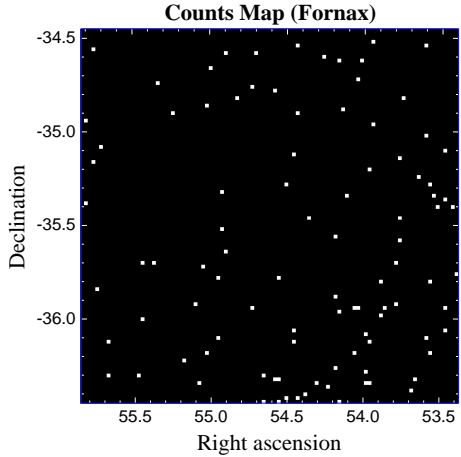}
\caption{The same as the left panel of Fig.~\ref{fig:maps_Fornax}, but
 with finer pixelization. White dots represent single photons received.}
\label{fig:counts_map_Fornax}
\end{center}
\end{figure}

\begin{figure}
\begin{center}
\includegraphics[width=7.0cm]{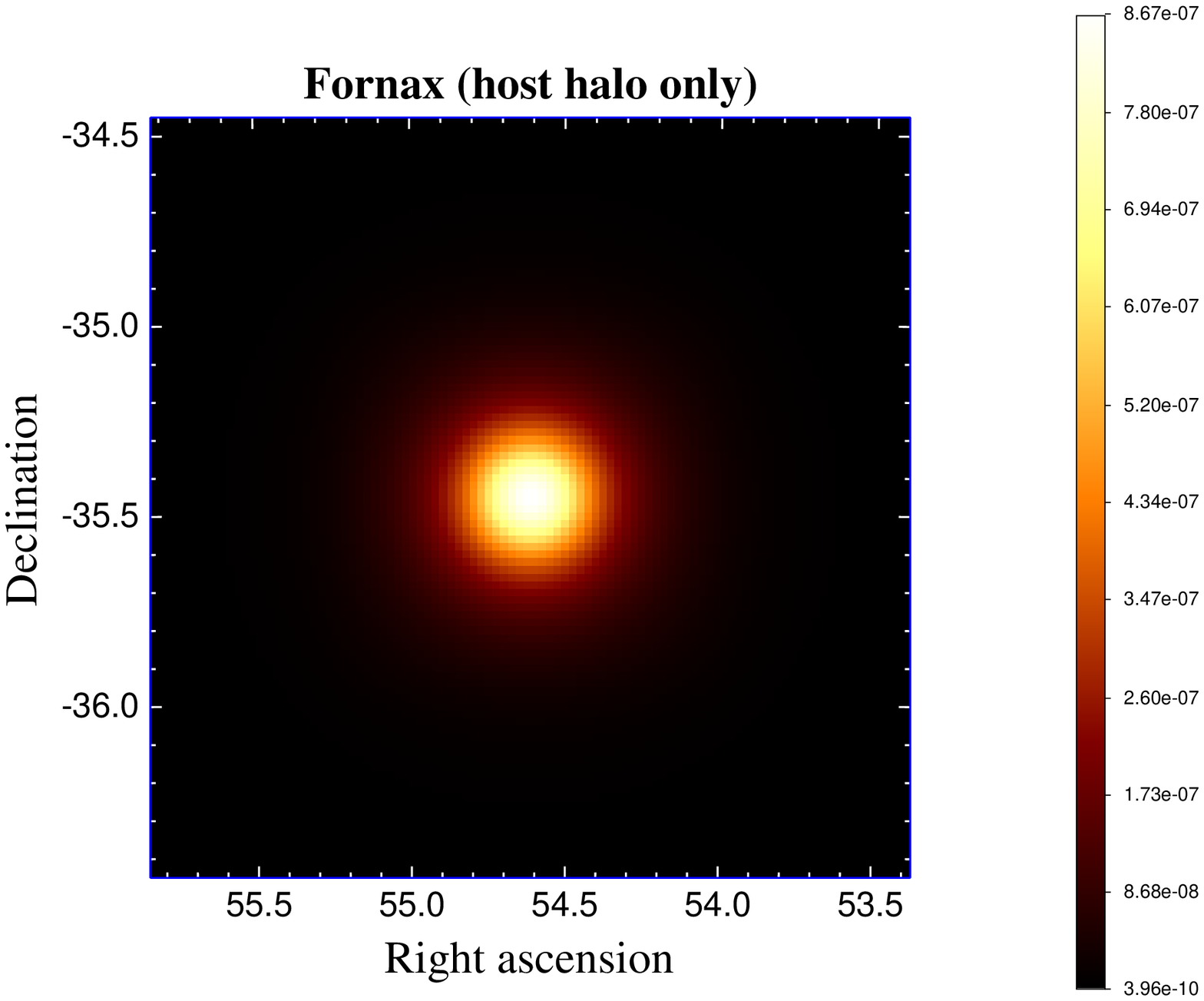}
\hspace{0.5cm}
\includegraphics[width=7.0cm]{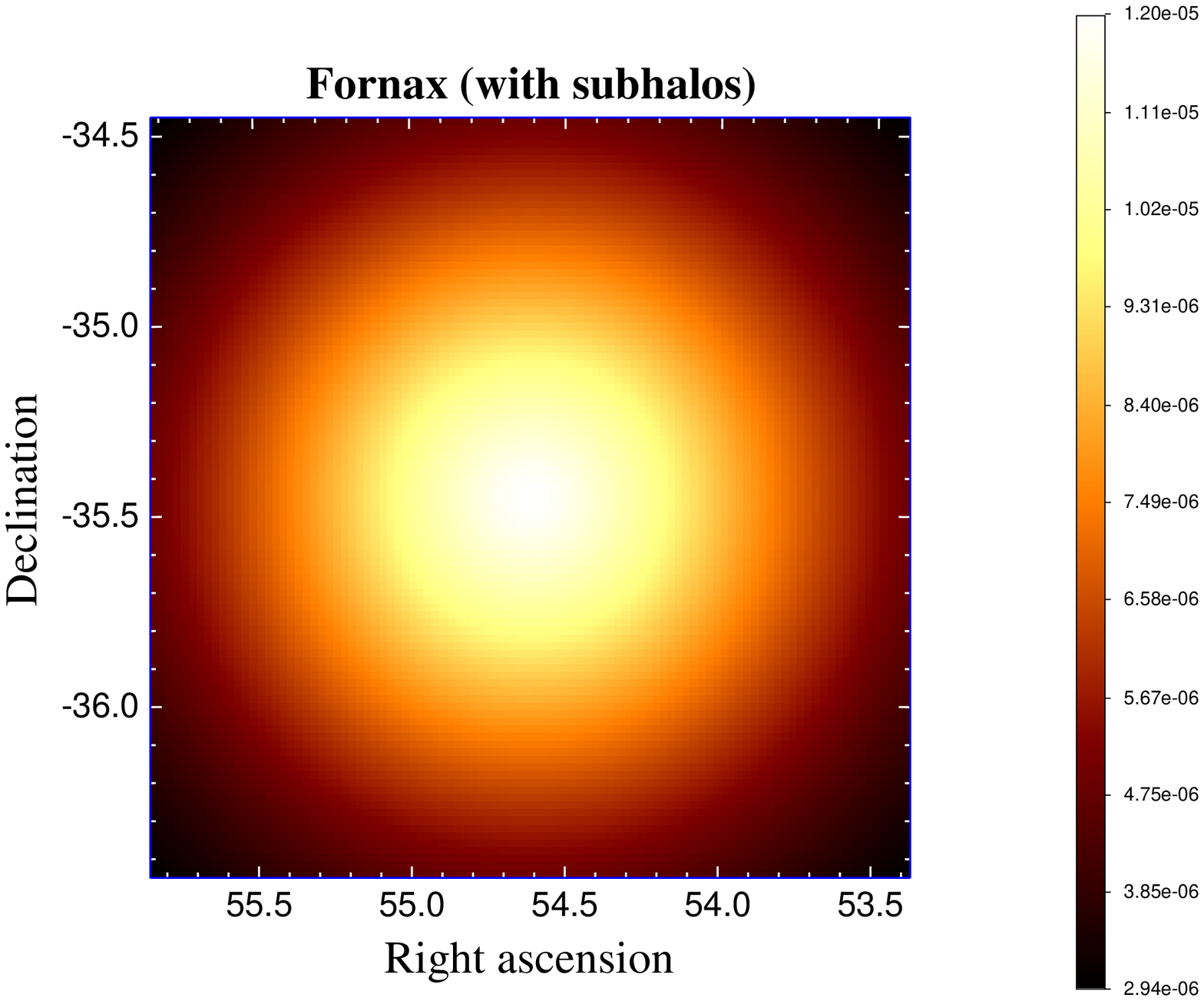}
\caption{The gamma-ray model maps of Fornax for the NFW density
 profile (left) and host plus subhalo model (right).  The
 colors represent expected photon counts per $0.02^\circ
 \times 0.02^\circ$ pixel (in linear scaling), where $b\bar b$
 annihilation channel with $m_\chi = 100$~GeV and $\langle \sigma v
 \rangle = 10^{-26} ~ \mathrm{cm^3 ~ s^{-1}}$ is assumed, and photon
 energies are between 7.94~GeV and 10~GeV. No backgrounds and other
 sources are included in these maps.}
\label{fig:Fornax}
\end{center}
\end{figure}

To make templates of cluster emission, we convolve cluster brightness
profiles with the P6\_V11 instrumental response functions (IRFs)
toward the direction of ROI.
Since the size of the cluster (i.e., $r_s / d_A$) and the point spread
function (PSF) is much smaller than ROI, we here use smaller regions
($\sim 2^\circ \times 2^\circ$) with finer pixel sizes ($0.02^\circ
\times 0.02^\circ$) around clusters for obtaining the upper limits.
For this new pixelization, we again run {\it gtbin} to make photon count
maps such as the one shown in Fig.~\ref{fig:counts_map_Fornax}, and {\it
gtmodel} to make model maps of the cluster as well as those of the
backgrounds and the other sources.
We show the model maps for the Fornax cluster after convolving with the
IRFs in Fig.~\ref{fig:Fornax} for both smooth
NFW host-halo case and the case with large boost due to subhalos (with
$M_{\rm min} = 10^{-6} M_\odot$).
One can see that the latter is much more extended than the former.
Note, however, that the backgrounds and other sources are not included
in these maps.
Although we do not show the maps for the contracted cluster model, they
look similar to the map for NFW.

\begin{figure}
\begin{center}
\includegraphics[width=8cm]{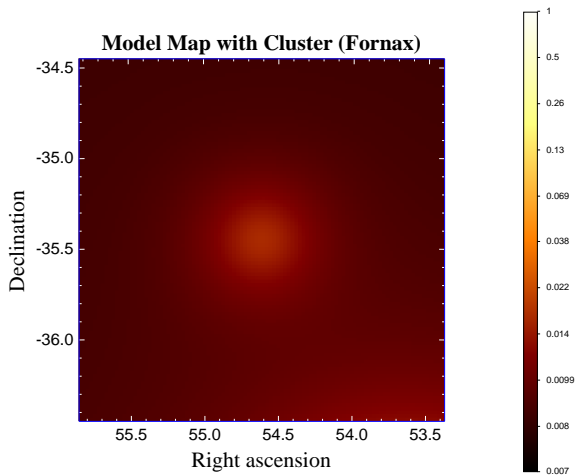}
\caption{Model map of the Fornax cluster combined with the diffuse
 backgrounds as well as point sources, integrated for 1--100~GeV. This
 map is to be compared with the counts map of
 Fig.~\ref{fig:counts_map_Fornax}; the color scaling is set such that
 white represents single photon received per pixel, while black does
 less than 0.007 photon.  The cluster dark matter component is from the
 smooth NFW halo, with $b\bar b$ annihilation channel, $m_\chi =
 100$\,GeV, and $\langle \sigma v \rangle = 5.5 \times 10^{-24} \,
 \mathrm{cm^3 \, s^{-1}}$ that is the 95\% credible upper limit from the
 current data.}
\label{fig:cluster_map_Fornax}
\end{center}
\end{figure}

We obtain upper limits on cluster emission by comparing the models with
the data.
According to the Bayesian statistics (e.g., \cite{Trotta2008}), the
posterior probability distribution function for some theoretical
parameter $\theta$ (in this case, the annihilation cross section,
$\theta = \langle \sigma v \rangle$) given data $d$ is
\begin{equation}
 P(\theta | d) \propto P(\theta) L(d | \theta),
\end{equation}
where $P(\theta)$ is the prior distribution, for which we adopt the
uniform and improper prior, $P(\theta) = {\rm const.}$, for positive
$\theta$ and zero otherwise, and $L(d | \theta)$ is the Poisson
likelihood function.
The latter is specifically given as
\begin{equation}
 L = \prod_{i=1}^{N_{\rm pix}} \frac{\mu_i^{d_i} e^{-\mu_i}}{d_i!},
  \label{eq:likelihood}
\end{equation}
where subscript $i$ represents pixel number, $\mu_i$ and $d_i$ are
theoretical model and the data counts in the pixel $i$, respectively.
We assume that the model is divided into the signal ($s_i$) and
background ($b_i$) contributions, $\mu_i = s_i + b_i$, where the former
depends on the theoretical parameter $\theta$ and the latter includes both
the diffuse backgrounds and 2FGL sources.
We here fix all the parameters for $b_i$ to the values obtained with
{\it gtlike} in the previous subsection, and then obtain an upper limit on
$\theta$ by solving
\begin{equation}
 1-\alpha = \int_0^{\theta_{\rm lim}} d\theta\ P(\theta | d),
\end{equation}
for $\theta_{\rm lim}$.
In particular, we are interested in upper limits corresponding to 95\%
credible interval, and for that we adopt $\alpha = 0.05$ above.
As an example, in Fig.~\ref{fig:cluster_map_Fornax}, we show what the
Fornax cluster map would appear when it is just as bright as currently
allowed and is dominated by the smooth NFW component (with no
contraction).
The photon counts per $0.02^\circ \times 0.02^\circ$ pixel could be at
most 0.017 at the center of the map.

When we obtain upper limits with stacked cluster maps, we generalize
Eq.~(\ref{eq:likelihood}) to include contributions from different
clusters as
\begin{equation}
 L = \prod_{cl,i} \frac{\mu_{cl,i}^{d_{cl,i}} e^{-\mu_{cl,i}}}{d_{cl,i}!},
  \label{eq:likelihood stack}
\end{equation}
where the subscript $cl$ represents clusters.
Here we remove some clusters from the stacking analysis, if they are
brighter than $3\sigma$ level compared with the backgrounds for any dark
matter masses and the annihilation channels.
For this purpose, we also use the posterior probability distribution
function introduced above, $P(\theta | d)$, and by fixing the parameters
for the background models $b_i$.
As the result, we have 38 clusters used in the stacking analysis.
A more careful treatment for the gamma-ray detection from the galaxy
clusters should be made by varying background parameters together with
$\theta$.
However, this is beyond the scope of the present study and will be
revisited in future publication.

\end{document}